	\newcommand{\VK}{\mbox{${\rm (V-K)}$}}

	\newcommand{\UVex}{\mbox{${\rm (1550-V)}$}}

	\newcommand{\Msun}{\mbox{${\rm M_{\odot}}$}}

        \newcommand{\dydz}{\mbox{${\rm \Delta Y/ \Delta Z}$}}

        \newcommand{\ML}{\mbox{${\rm M_{L}}$}}

\def\smallskip{\vskip 8pt}
\def\littleskip{\vskip 6pt}
\def\verylittleskip{\vskip 2pt}

\def\M12{${\rm M_{L,12}} $}

       \documentstyle[]{l-aa}
        \input psfig.tex

\begin{document}

\thesaurus{       }

\title{Spectro-photometric Evolution of Elliptical Galaxies. II. }

\subtitle{Models with Infall }

\thanks{  }
 
\author {R. Tantalo$^1$, C. Chiosi$^1$, A. Bressan$^2$,  F. Fagotto$^{3,1}$  }

\institute {
$^1$ Department of Astronomy, Vicolo dell' Osservatorio 5, 35122 Padua, 
   Italy\\
$^2$ Astronomical Observatory, Vicolo dell' Osservatorio 5, 35122 Padua, 
   Italy\\   
$^3$ Instituto de Astrofisica de Canarias, 38200 La Laguna, Tenerife, 
Spain }

\offprints{C. Chiosi }

\date {Received July 1995, accepted }

\maketitle

\markboth{Tantalo et al.: Evolution of Elliptical Galaxies. II. Models with 
Infall }{}

\begin{abstract}
In this paper we present new chemo-spectro-photometric models of elliptical
galaxies in which infall of primordial gas is allowed to occur. They 
aim to simulate the collapse of a galaxy made of two components, i.e. 
luminous material and dark matter. The mass of the dark
component is assumed to be constant in time, whereas that of the luminous
material is supposed to accrete at a suitable rate. They also include the
effect of galactic winds powered by supernova explosions and stellar winds from
massive, early-type stars. The models are constrained to match a
number of properties of elliptical galaxies, i.e.  the slope and mean colours
of the colour-magnitude relation (CMR), V versus (V--K),
the UV excess as measured by the colour (1550--V) together with the overall
shape of the integrated spectral energy distribution (ISED) in the 
ultraviolet,  the relation between the ${\rm Mg_2 }$ index and (1550--V), 
the mass to blue luminosity ratio 
${\rm M/L_{B}}$ as a function of the ${\rm B}$ luminosity, and finally the
broad-band colours (U--B), (B--V), (V--I), (V--K), etc. 
The CMR is interpreted as a mass-metallicity sequence of old, nearly coeval
objects, whose mean age is 15 Gyr.
Assuming the law of star formation to be proportional 
to ${\rm M_{g}^{k}(t)}$ with $k = 1$, the rate of
star formation as function of time starts small, grows to a maximum, and then
declines thus easily avoiding the excess of metal-poor stars found by BCF with
the closed-box scheme (the analog of the G-Dwarf Problem in the solar
vicinity). Owing to their stellar content, infall models can easily reproduce
all the basic data of the galaxies under examination. As far as the UV excess
is concerned, the same sources proposed by BCF are found to hold also with the
infall scheme. H-HB and AGB manqu\'e stars of high metallicity play the
dominant role, and provide a robust explanation of the correlation
between the  (1550--V) colour and the luminosity, mass
and metallicity of the galaxies.  Furthermore, these models confirm the
potential  of the (1550--V) colour as an  age indicator in
cosmology as already suggested by BCF. In the rest frame of a massive 
and metal-rich elliptical galaxy, this
colour suffers from one major variation: at the onset of the so-called H-HB
and AGB-manqu\'e stars (age about 5.6 Gyr). This transition occurs at 
 reasonably small red-shifts and therefore could be detected with 
the present-day instrumentation. 


\end{abstract}

\section{Introduction}
Bressan et al. (1994,\ BCF) have recently presented 
chemo-spectro-photometric models for elliptical galaxies 
that were  particularly designed to match the
UV excess observed in these systems  (Burstein et al.\ 1988) together with
its dependence on the galaxy luminosity, mass, metallicity, ${\rm Mg_{2}}$ 
index,  and age. 
In brief, BCF models stand on the closed-box approximation and 
the enrichment law 
$\Delta Y/\Delta Z = 2.5$,  allow for the occurrence of galactic winds
powered by the energy input from supernovae and stellar winds from massive 
stars, and use  the CMR of elliptical galaxies in Virgo and Coma clusters 
(Bower et al. 1992a,b) as one of the main constraints. 

In the BCF models the 
history of star formation consists of an initial period of activity 
followed by quiescence  after the onset of the galactic winds, whose duration 
depends on the galactic mass, being longer in the high mass galaxies and 
shorter in the low mass ones. With the adoption of the closed-box description 
of chemical evolution, the maximum efficiency of star formation occurred at the 
very initial stage leading to a  metallicity distribution 
in the models (relative number of stars per metallicity interval)
 skewed towards low metallicities.

The comparison of the theoretical integrated spectral energy distribution
(ISED) of the models with those of prototype galaxies with different intensity
of the UV excess, for instance NGC~4649 with strong UV emission and 
(1550--V) $\sim
2.24$  and NGC~1404 with intermediate UV excess and (1550--V) $\sim 3.30$,
indicated that a good agreement was possible but for the region 2000 $\AA$ to
about 3500 $\AA$ where the theoretical ISED exceeded the observed one.

%%%%%%%%Figure 1a %%%%%%%%%%%
\begin{figure}
%\picplace{9cm}
\psfig{file=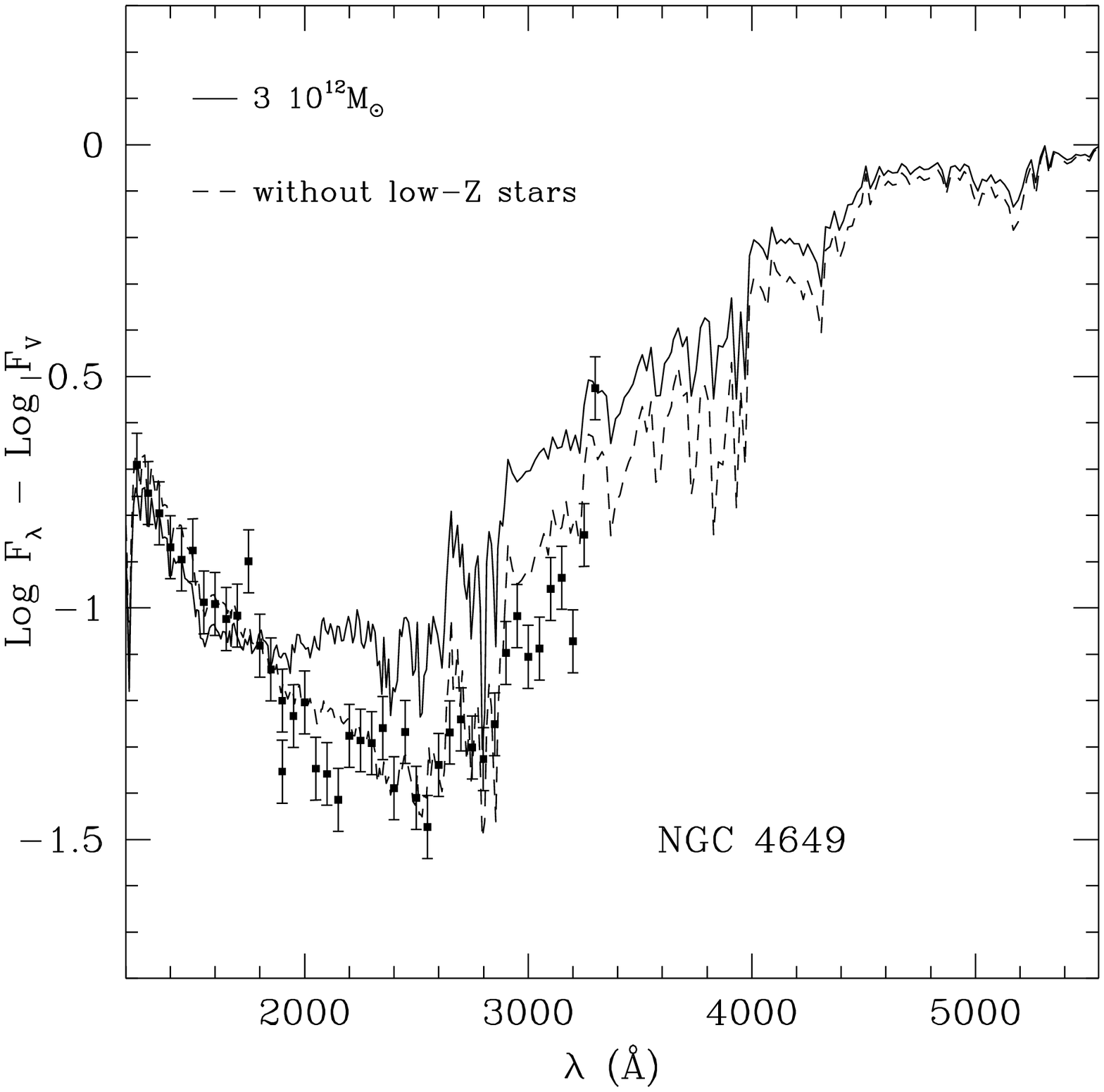,height=9.0truecm,width=8.5truecm}
\caption{The observed spectrum of the elliptical galaxy NGC~4649 
(filled squares) kindly provided by L. Buson (1995, private communication).
The vertical bars show the global error on the (1550--V) colours from
Burstein et al. (1988).
The two  lines show theoretical spectra for the galactic model of BCF with
 \ML = ${\rm 3\time10^{12} M_{\odot} }$ characterized by the parameters
 ${\rm k=1}$, $\nu=20$ and $\zeta=0.40$ and age of 15 Gyr, obtained with
different assumptions concerning the contribution from stars in different
metallicity bins. The solid  line is the regular spectrum in which all the 
components are present. Note the discrepancy in the wavelength interval 
2000 - 4000 \AA.  The dashed 
line is the same but without the contribution from all stars with 
metallicity ${\rm Z \leq 0.008}$. Note the agreement reached in this case.
All the spectra are normalized to coincide in the flux at 5000 \AA.  } 
\label{fig1a}
\end{figure}

The analysis of the reason of disagreement led  BCF to
conclude that this is a consequence of the closed-box approximation. Indeed
this type of model is known to predict an excess of low-metal stars. The
feature is intrinsic to the model and not to the particular numerical algorithm
 used to follow the chemical history of a galaxy. 

In order to check their suggestion, BCF performed several numerical experiments
in which they artificially removed from the mix of stellar populations
predicted by the closed-box models the contribution to the ISED from stars in
different metallicity bins. Looking at the paradigmatic case of the ISED of
NGC~4649, BCF found that this can be matched by a mix of stellar populations in
which no stars with metallicity lower than ${\rm Z=0.008 }$ are present. 
The results of these experiments are shown here in Fig.~\ref{fig1a} for the 
sake of clarity. 

It goes without saying that one may 
attribute the discrepancy of predicted and observed spectra in the region 
2000 to 3500 \AA\ to uncertainties in the theoretical spectra. However, the 
following considerations can be made. Firstly, those 
 experiments clarified that the main 
contributors to the excess of flux are the turn-off stars of low 
metallicity ($\rm 0.0004 \leq Z \leq 0.008 $)  whose effective temperature 
at the canonical age of 15 Gyr ranges from 6450 K to 5780 K. 
Secondly, the  Kurucz (1992) spectra (at the base of the library adopted 
by BCF) 
fit the Sun and Vega, two stars of different metallicity but 
whose effective temperatures encompass the above values.
Finally,  the 
disagreement in question implies  that the theoretical spectra 
overestimate the flux in this region by a factor of about 4, which is hard 
to accept. On the basis of the above arguments  
BCF concluded that the excess is real and that the discrepancy in question 
is the analog
of the classical G-Dwarf Problem in the solar vicinity. 

In brief, it has long been known that the closed-box model applied to study the
chemical history of the solar vicinity fails to explain the metallicity
distribution observed among the old field stars, giving rise to the so-called
G-Dwarf Problem  (Tinsley 1980a,b). The solution of the G-Dwarf dilemma are
models with infall (Lynden-Bell 1975, Tinsley 1980a, Chiosi 1981), i.e. models
in which the total mass of the disk is let increase with time at a convenient
rate starting from a  much lower value. With the current law of star formation
(proportional to the gas mass) the competition between gas accretion by infall
and gas depletion by star formation gives rise to a non monotonic time
dependence of the star formation rate which instead of steadily decreasing from
the initial stage as in the closed-box model starts small, increases to a peak
value, and then declines over a time scale which is a sizable fraction of the
infall time scale. The advantage of the infall model with respect to the
closed-box one, is that the metallicity increases faster, and very few stars
are formed at very low metallicity. Since the excess of very low-metal stars is
avoided, the G-Dwarf Problem is naturally solved. 

Applied to an elliptical galaxy, the infall model of chemical evolution can
closely mimic the collapse of the parental proto-galaxy from a very extended
size to the one we see today. Plausibly, as the gas falls into the
gravitational potential well at a suitable rate and the galaxy shrinks, the gas
density increases so that star formation begins. As more gas flows in, the more
efficient star formation gets. Eventually, the whole gas content is exhausted
and turned into stars, thus quenching further star formation. Like in the disk,
the star formation rate starts small, rises to a maximum, and then declines.
Because of the more efficient chemical enrichment of the infall model, the
initial metallicity for the bulk of star forming activity is significantly
different from zero. 
 
In this paper, we present simple models of spectro-photometric evolution of
elliptical galaxies whose chemical enrichment is governed by the infall scheme,
i.e. models in which the mass of the galaxy is let increase with time. Models
of this type are consistent with the chemo-dynamical models by Theis et al.\
(1992) and are advocated by Phillips (1993) to interpret the galaxy counts at
faint magnitudes, which are known to exceed those expected from standard
evolutionary models with an initial spike of star formation followed by
quiescence. 

As in BCF, the models (both chemical and spectro-photometric) are constrained
to reproduce at the same time a number of key features of elliptical galaxies,
namely the slope and mean colours of the CMR, V versus (V--K), 
 for the galaxies in the Virgo and Coma clusters,
the UV excess as measured by the colour (1550--V) together with the overall
ISED, the relation between the ${\rm Mg_2 }$ index and (1550--V), 
the mass to blue
luminosity ratio ${\rm M/L_B }$ as  a function of the ${\rm B}$ 
luminosity, and finally the
present broad band colours (U--B), (B--V), (V--I), (V--K), etc. 

The CMR is from
Bower et al. (1992a,b), the data on the UV excess and ${\rm Mg_2 }$-(1550--V)
relationship are from Burstein et al.\ (1988), and the data on the ratio
${\rm M/L_B }$ 
are from Bender et al.\ (1992,1993) and Terlevich \& Boyle (1993) for
${ \rm H_0=50~ km~ sec^{-1}~ Mpc^{-1}  }$. 

There is one aspect of the problem that should be clarified in advance. 
The bottom line of BCF models is the interpretation 
of the Virgo and Coma cluster CMR as a mass-metallicity sequence  
whose slope is driven by the onset of galactic winds (cf. Larson 1974; 
Saito 1979a,b; Vader 1986; Arimoto \& Yoshii 1986; Matteucci \& Tornamb\'e
1987; Angeletti \& Giannone 1990; Padovani \& Matteucci 1993) and whose 
tightness  is the signature of old ages with small dispersion, 
$13 \div 15$ Gyr according to Bower et al. (1992a,b). 

In contrast, the CMR for nearby galaxies in small groups and field is more 
dispersed than the one for the Virgo and Coma clusters suggesting that 
elliptical galaxies may originate
 from the mergers of spiral galaxies spread over 
several billion years (Schweizer \& Seitzer 1992; Alfensleben \& Gerhard 
1994; Charlot \& Silk 1994). 
Also in this case the CMR implies the existence of 
a suitable mass-metallicity sequence. 

Therefore, the view that elliptical 
galaxies are primordial old objects with a dominant 
 initial burst of activity may co-exist
 with the view that many if not all of them are the result  
of mergers  implying  more recent star formation 
activities that may vary from galaxy to galaxy.

How the  different interpretation of the CMR   affects the use of this 
as a  constraint on our models ? This question has been addressed by 
Bressan et al. (1995, \ BCT), who showed that the slope and the mean 
colours of the CMR are compatible with both alternatives,  thus validating 
its ability to  constrain the models. Therefore, owing to its better 
definition, we adopt the Bower et al. (1992a,b) CMR and assume that it
is made of  nearly coeval objects whit  mean age of about 15 Gyr.
Strictly speaking the  models below are best suited to cluster
elliptical galaxies. The analysis of the merger scenario  is beyond the 
scope of this study.

Section 2  reports on the new spectral library adopted to calculate the
spectro-photometric models presented in this paper. Section 3 describes the
properties of Single Stellar Populations (SSP) calculated with the new spectral
library above. Section 4 outlines the chemical models in usage. Section 5
contains closed-box models calculated with the new spectral library and
compares them with those of BCF. It is found that no closed-box model can match
all the imposed constraints. Section 6 is the same but for infall models.
Section 7 presents the results of the chemical and spectro-photometric models
with infall and describes in detail the properties of the models with different
mass. In particular we examine the CMR, the mass to blue luminosity ratio
${\rm M/L_{B} }$, 
the UV excess and accompanying (1550--V) versus ${\rm Mg_{2}}$ relation,
the evolution of the broad-band colours as a function of the age, and the
potential capability of the (1550--V) colour of being a good age indicator in
cosmology. Finally,  Section 8 contains a few concluding remarks.

\section{ A new spectral library }\label{spec_lib}

The library of stellar spectra adopted in this study is very similar to the 
one used by BCF. There are however a few important changes  deserving 
a bit of explanation.

The main body of the spectral library is from R. Kurucz (1992),
however extended to  the 
high and low temperature ranges.
For stars with high ${\rm T_{eff}} >  50,000$ K pure
black-body spectra are assigned, whereas for stars with ${\rm T_{eff}} < 3500$
 K the new catalog of stellar fluxes by 
Fluks et al. (1994) is adopted (see Silva 1995 for details). This
 catalog includes 97
observed spectra for all M-spectral subtypes in the wavelength range $3800 \leq
\AA \leq 9000$, and synthetic photospheric spectra in the range $9900 \leq \AA
\leq 12500$. 

The scale of  ${\rm T_{eff} }$ in Fluks et al. (1994) is similar
to that of Ridgway et al. (1980) for spectral types earlier than M4 but
deviates from it for later spectral types. Since Ridgway's et al. (1980)
scale does not go beyond the spectral type M6, no comparison for more
advanced spectral types is  possible. 

The problem is further complicated by possible effects of metallicity. The
Ridgway  scale of ${\rm T_{eff}}$ is based on stars with solar metallicity 
(${\rm Z \sim 0.02}$) and empirical  calibrations of the  ${\rm T_{eff}}$-scale 
for ${\rm Z \neq 0.02}$ are not available. In contrast, the library of
SSPs span the range of metallicity ${\rm 0.0004 \leq Z \leq 0.1 }$. 

To cope with this difficulty, we have introduced the 
metallicity-${\rm T_{eff}}$ relation of Bessell et al.
(1989,1991) using the (V--K) colour as a temperature
indicator. An interpolation is made between the ${\rm T_{eff}}$ of 
Bessell et al.\ (1989)
and the (V--K) colours given by Fluks et al. (1994) for the spectral
types from M0 to M10. 

\begin{figure}
%\picplace{9cm}
\psfig{file=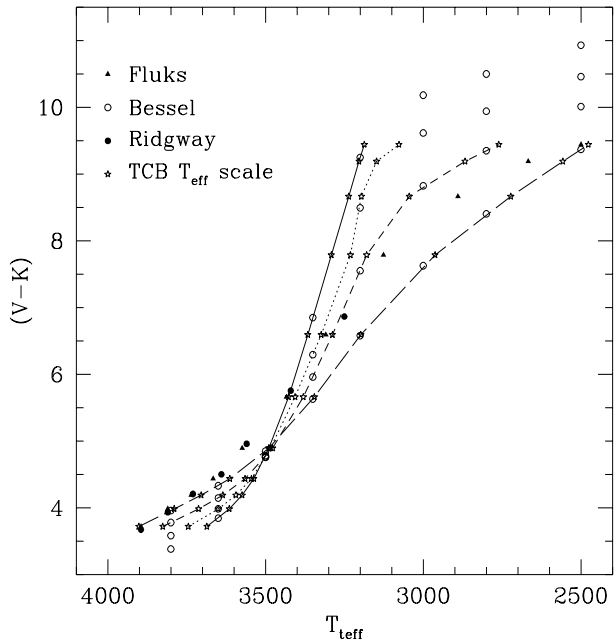,height=9.0truecm,width=8.5truecm}
\caption{ Four (V--K) versus ${\rm T_{eff}}$ scales:
Ridgway et al. (1980)
--- full dots, Bessell et al. (1989, 1991) ---  open dots, Fluks et al.
(1994) for solar metallicity --- solid triangles. Finally, the  asterisks
correspond to  our data. The lines show the interpolations between the (V--K)
colours of Fluks et al. (1994)  and the ${\rm T_{eff}}$ scales of Bessell et al.
(1989, 1991) for different metallicities } 
\label{fig1}
\end{figure}

Figure~\ref{fig1} compares the ${\rm T_{eff}}$  scales of Ridgway et al. 
(1980),
Fluks et al. (1994) and Bessell et al. (1989, 1991) with the one we have
obtained. This latter is also given in Table~1 as a function of the spectral
type and metallicity. Column (1) is the spectral type, column (2) is the (V--K)
colour from Fluks et al. (1994) for solar metallicity, column (3) through (6)
are the ${\rm T_{eff}}$ for different metallicities (Z=0.06, Z=0.02, Z=0.006,
Z=0.002) as indicated.

%%%%%%%%Figure 1 %%%%%%%%%%%

%%%%%%%%%Table 1 (la scala di temperature calcolata da Silva) %%%%%
\begin{table}[htb]
\vskip 0.3 cm
\begin{center}
\vskip 0.2 cm
\caption{ Spectral type, (V--K) colour, and ${\rm T_{eff}}$ scales as a 
function of the metallicity}
\vskip 0.25 cm
\scriptsize
\begin{tabular} {c c c c c c}
\hline
\hline
 & & & & & \\
SpT & (V--K) & $T_{0.06}$ & $T_{0.02}$ & $T_{0.006}$ & $T_{0.002}$ \\
 & & & & & \\
\hline
 & & & & & \\
M0  & 3.722 & 3686 & 3745 & 3825 & 3902 \\
M1  & 3.986 & 3615 & 3650 & 3712 & 3789 \\
M2  & 4.190 & 3574 & 3594 & 3636 & 3704 \\
M3  & 4.439 & 3537 & 3545 & 3565 & 3613 \\
M4  & 4.895 & 3489 & 3484 & 3477 & 3490 \\
M5  & 5.662 & 3427 & 3406 & 3380 & 3345 \\
M6  & 6.595 & 3366 & 3324 & 3288 & 3198 \\
M7  & 7.789 & 3291 & 3231 & 3179 & 2963 \\
M8  & 8.667 & 3236 & 3196 & 3044 & 2722 \\
M9  & 9.192 & 3204 & 3148 & 2868 & 2558 \\
M10 & 9.444 & 3187 & 3077 & 2760 & 2476 \\
 & & & & & \\
\hline
\hline
\end{tabular}
\end{center}
\label{tab1}
\end{table}

\section{ A new library of SSPs with $\Delta Y/\Delta Z$=2.5 }\label{ssp_lib}

BCF presented an extended library of integrated magnitudes and colours of SSPs
as a function of the age for several combinations of the chemical parameters Y
(helium) and Z (metallicity) obeying the enrichment law $\Delta Y/\Delta Z=2.5$
(Pagel 1989, Pagel et al. 1992). The sets of chemical abundances were [Y=0.230,
Z=0.0004], [Y=0.240, Z=0.004], [Y=0.250, Z=0.008], [Y=0.280, Z=0.020],
[Y=0.352, Z=0.05], and [Y=0.475, Z=0.1]. 

The stellar models used to calculated the SSPs with $\Delta Y/\Delta Z=2.5$ in
BCF were from Alongi et al. (1993), Bressan et al. (1993) and Fagotto et al.
(1994a,b,c). They  span the mass range from 0.6 to 
${\rm 120 M_{\odot} }$
and go from the zero age main sequence until the start of the thermally pulsing
asymptotic giant branch stage (TP-AGB) or C-ignition as appropriate to  the
intial mass of the star. 

In this paper we adopt the same library of stellar models to calculate our SSPs
however with two major differences with respect to BCF. First we extend the
zero age main sequence down to ${\rm 0.15 M_{\odot}}$ 
using the VanderBerg (1985)
library of low mass stars (the lowest mass in BCF was 
${\rm 0.6 M_{\odot} }$). Second
we  increase the mass loss parameter $\eta$ governing stellar winds from RGB
and AGB stars from $\eta=0.35$ to $\eta$=0.45 (see BCF for more details). 

The method of SSP calculation is exactly the same as in BCF, to whom we refer
for all details. Suffice it to recall that SSP models are for the IMF expressed
by equations (\ref{imf}) below, with slope $x=2.35$ and
lower and upper limits of integration, ${\rm M_L = 0.15  M_{\odot}}$
and ${\rm M_U = 120  M_{\odot}}$ respectively. It is worth recalling 
that the integrated magnitudes of
these ideal SSPs  must be properly scaled when applied to real SSPs (for
instance star clusters). 

Table 2 contains the integrated magnitudes and colours of the new SSPs for
different values of the age. We list the age in Gyr (column 1), the absolute
visual magnitude ${\rm M_V}$ (column 2), the absolute bolometric magnitude 
${\rm M_{bol} }$
(column 3), the bolometric corrections BC (column 4), and the colours (U--B),
(B--V), (V--R), (R--I), (V--J), (V--K), (V--L), (V--M), (V--N), and (1550--V),
columns (5) through (14), respectively.

%%%%%%%%%Table 2 (Integrated colours of SSPs) %%%%%

\begin{figure}
%\picplace{ 14cm}  
\psfig{file=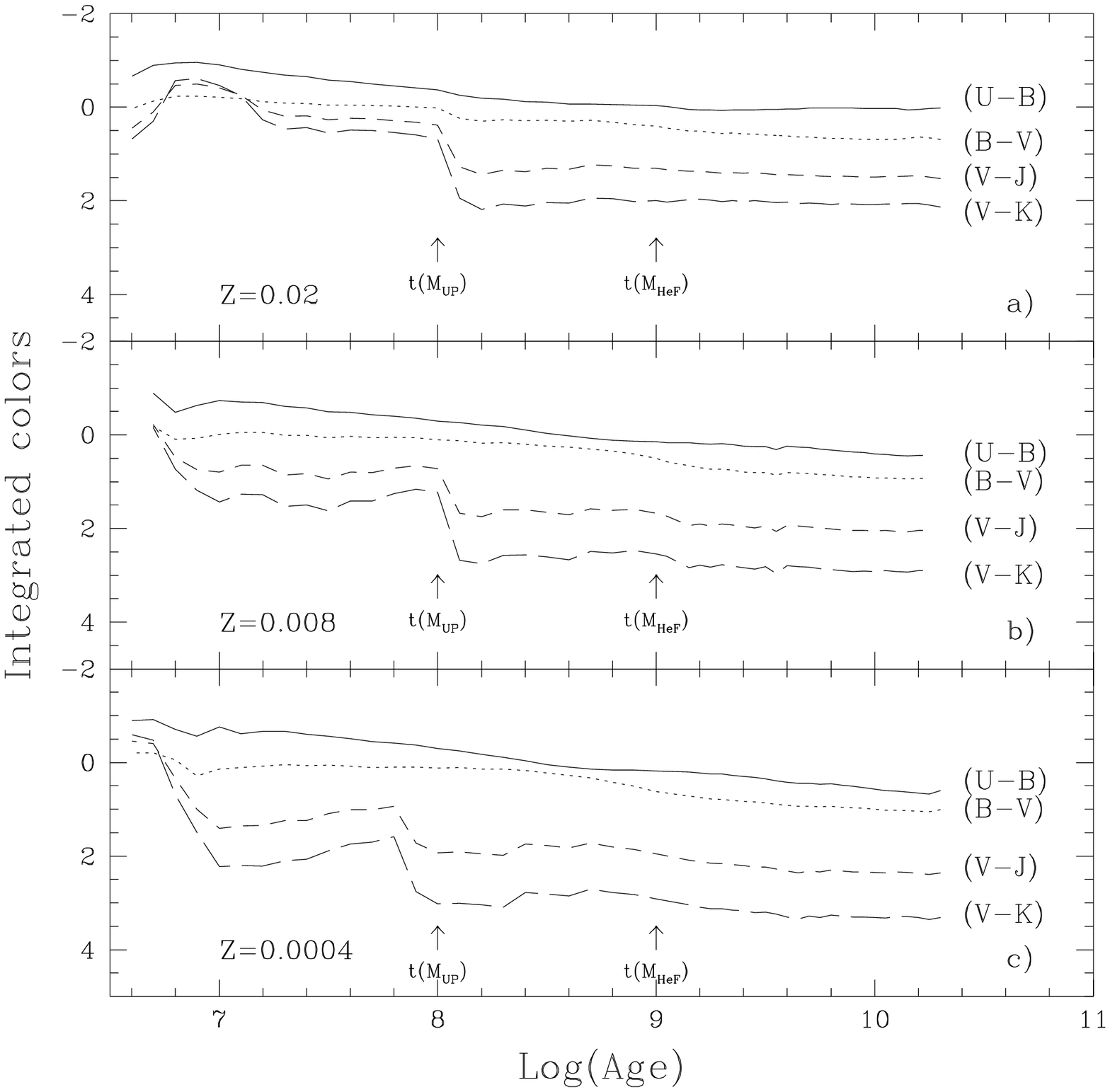,height=14.0truecm,width=8.5truecm}
\caption{{\bf Panel (a): } The colour evolution of a SSP with composition
[Y=0.28 Z=0.02]. Note the large variation in the red colours caused by the
onset of AGB stars at the age ${\rm t(M_{UP}}$. The appearance of RGB stars 
expected to occur at the age ${\rm t(M_{HeF}}$
is masked by
AGB stars so that no change in the colour is noticed. Ages are in years. {\bf
Panel (b): }  The same as in Panel  (a) but for a SSP with composition [Y=0.25
Z=0.008]. {\bf Panel (c): } The same as in Panel (a) but for a SSP with
composition [Y=0.23 Z=0.0004] } 
\label{fig2}
\end{figure}

Figure~\ref{fig2} shows the temporal evolution of four selected colours, namely
(U--B), (B--V), (V--J) and (V--K) for three SSPs with different chemical
composition: Panel (a) for [Y=0.28 Z=0.02], Panel (b) for [Y=0.25 Z=0.008], 
and Panel (c) for [Y=0.23 Z=0.0004]. The vertical arrows show the ages at which
a sudden variation in the colours caused by the onset of AGB and RGB stars is
expected to occur according to Renzini \& Buzzoni (1986). 

\begin{figure}
%\picplace{ 9cm}
\psfig{file=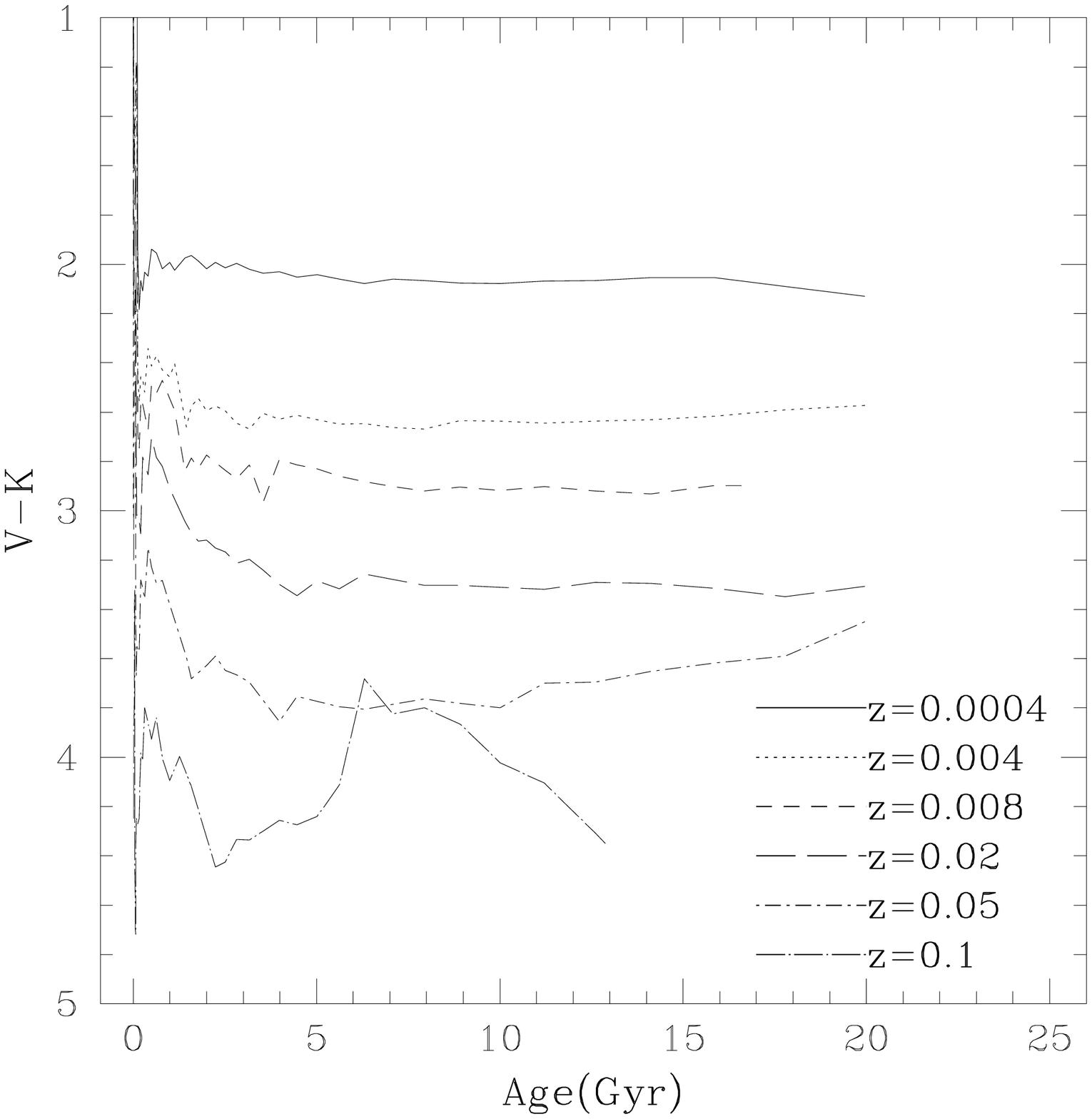,height=9.0truecm,width=8.5truecm}
\caption{ The temporal evolution of the (V--K) colour for SSPs with different
chemical compositions obeying the enrichment law \dydz = 2.5} 
\label{fig3}
\end{figure}

\begin{figure}
%\picplace{ 9cm}
\psfig{file=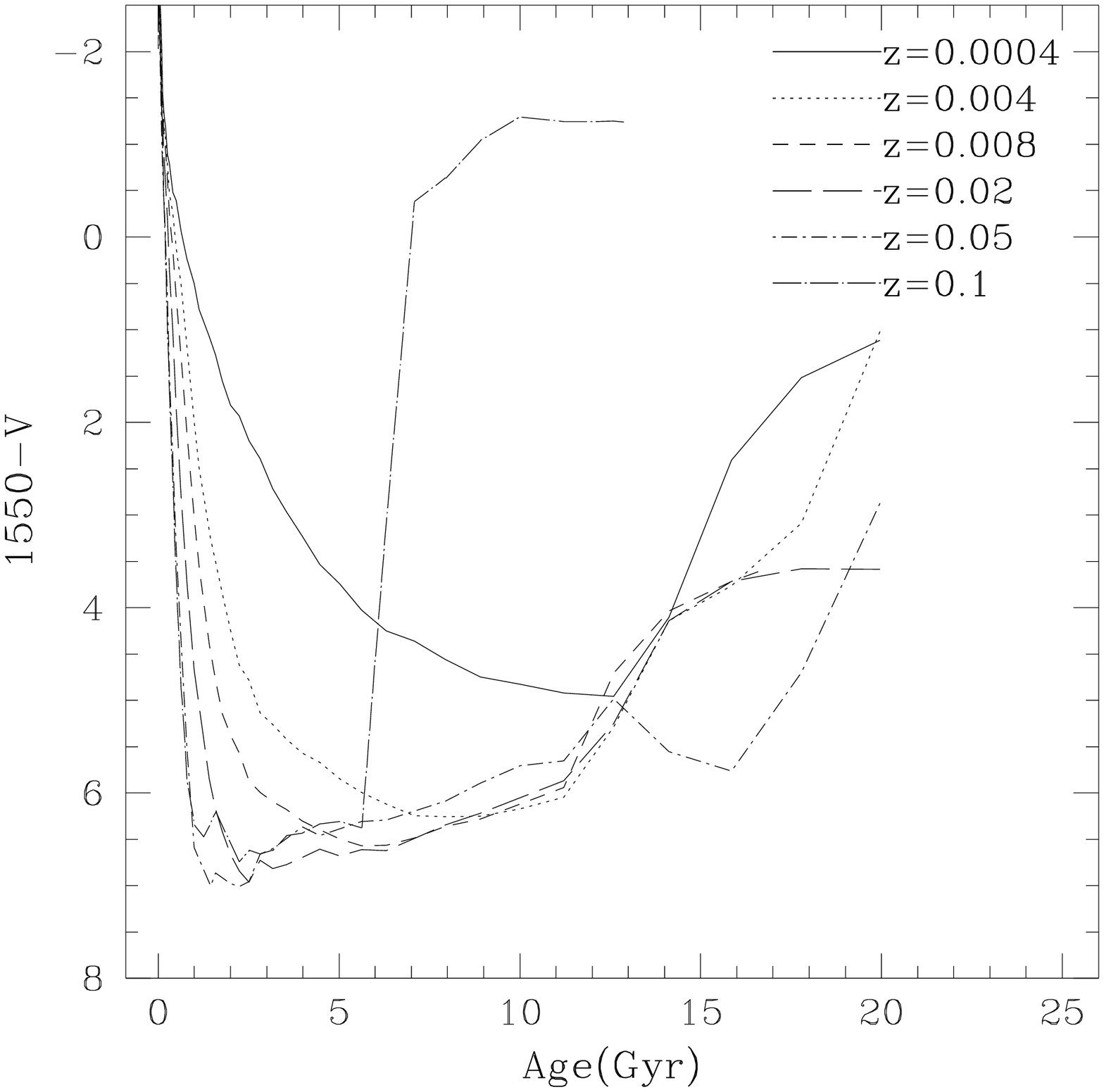,height=9.0truecm,width=8.5truecm}
\caption{ The temporal evolution of the (1550--V) colour for SSPs with 
different chemical compositions obeying the enrichment law \dydz = 2.5 } 
\label{fig4}
\end{figure}

Figures~\ref{fig3} and \ref{fig4} display the temporal evolution of the colours
(V--K) and (1550--V), respectively, for all the chemical compositions under
consideration.  While the  (V--K) colour remains almost constant for SSPs older
than 5~Gyr (cf.  Fig.\ref{fig3}), the (1550--V) colour (Fig.~\ref{fig4})
undergoes large variations. This implies that in order to match the observed
integrated (V--K) and (1550--V) colours, the galactic models (both closed and
open) must develop suitable mean and maximum metallicities (see below).

%%%%%%%%Figure 2 %%%%%%%%%%%

%%%%%%%%Figure 3 %%%%%%%%%%%

%%%%%%%%Figure 4 %%%%%%%%%%%

\section{Chemical models of elliptical galaxies }\label{ope_mod}

\subsection{Outline of the Model}\label{out_mod}
The model aims at simulating in a very simple fashion the formation of an
elliptical  galaxy by collapse of primordial gas in presence of dark matter. To
this scope, elliptical galaxies are described as one-zone models  made of two
components,  luminous material of mass ${\rm M_{L}(t) }$ 
 embedded in a halo of dark
matter of mass ${\rm M_{D} }$, 
whose presence affects only the gravitational potential
of the galaxy and the binding energy of the gas. 

In the infall scheme the mass of the luminous component (supposedly in form of
gas) is let increase with time according to 

%%%%%%%%%%%%equation 1
\begin{equation}
{\rm   {dM_{L}(t) \over dt} = \dot{M_{0}} ~exp(-t/\tau)  }  
\label{inf}
\end{equation}
where $\tau$ is the accretion time scale. 

The constant ${\rm \dot{M_{0}} }$ 
is obtained from imposing that at the galaxy age 
${\rm T_{G}}$ the value ${\rm M_{L}(T_{G})}$ is reached:

%%%%%%%%%%%%equation 2
\begin{equation}
{\rm   \dot{M_{0}} = \frac{M_{L}(T_{G})}{\tau 
[1 - exp(- T_{G}/\tau)]}   }  
\label{mdot}
\end{equation}

Therefore, integrating the accretion law the time dependence of 
${\rm M_{L}(t)}$ is 

%%%%%%%%%%%%equation 3
\begin{equation}
 {\rm M_{L}(t) =  { M_{L}(T_{G}) \over {[1-exp (-T_{G}/\tau)] } } 
                      [1 - exp(-t/\tau)]  }  
\label{mas-t}
\end{equation}

The spatial distribution of the dark component with respect to the luminous
one is supposed to obey the dynamical models of Bertin et al. (1992) 
and Saglia et al. (1992) in which 
the mass and radius of the dark component, ${\rm M_{D}}$ and ${\rm R_{D}}$,
respectively, are related to those of the luminous material,  ${\rm M_{L}}$ and
${\rm R_{L}}$, by the relation 

%%%%%%%%%%%%equation 5
\begin{equation}
{\rm  {M_{L}(t) \over M_{D} } \geq {1\over 2\pi} ({R_{L}(t)\over R_{D}}) 
       [1 + 1.37 ( {R_{L}(t) \over R_{D} } )]   }  
\label{dark}
\end{equation}

The mass of the dark component is assumed be constant in time
and equal to
${\rm M_{D}= \beta M_{L}(T_{G}) }$ with $\beta=5$.  

Furthermore, the binding gravitational energy of the gas is given by

%%%%%%%%%%%%%equation 6
\begin{equation}
{\rm \Omega_{g}(t)=-{\alpha}_{L} G {M_{g}(t) M_{L}(t)\over R_{L}(t) } - 
G {M_{g}(t) M_{D} \over  R_{L}(t) } \Omega'_{LD}  }   
\label{gas_pot}
\end{equation}
\noindent
 where ${\rm M_g(t)}$ is the current value of the gas
mass, $\alpha_L$ is numerical factor $\simeq 0.5$, and 

%%%%%%%%%%%%%equation 7
\begin{equation}
{ \rm \Omega'_{LD}= {1\over 2\pi} ({R_{L}(t)\over R_{D}}) [1 + 1.37
            ( {R_{L}(t) \over R_{D} } )] }  
\label{dark_pot}
\end{equation}
\noindent
is the contribution to the gravitational energy given by the presence of dark
matter. 
Following Bertin et al. (1992) and Saglia et al. (1992) in the above relations
we assume ${\rm M_{L}/M_{D}}=0.2$ and ${\rm R_{L}/R_{D}}=0.2$.

In order to apply the above equations for the binding gravitational energy of 
the gas to a model galaxy we need a relation between
the radius and the mass of the luminous component. To this aim,
we start noticing 
 that with the assumed ratios ${\rm M_{L}/M_{D}}$ and
${\rm R_{L}/R_{D}}$ the  contribution to gravitational energy by the 
dark matter is ${\rm \Omega'_{LD}=0.04}$ and the  total correction to the
gravitational energy of the gas (eq. \ref{gas_pot}) does not
exceed 0.3 of the term for the luminous mass.

Given these premises, and  assuming  
that at each stage of the infall process the amount of
luminous mass that has already accumulated gets soon virialized and turned 
into stars, we can approximate 
 the total gravitational energy and radius of the material
already settled onto the equilibrium configuration with the relations 
for the total
gravitational energy and radius as function of the mass (in this case
${\rm M_{L}(t)}$) obtained by Saito (1979a,b) for elliptical galaxies 
whose spatial distribution
of  stars is such that the global luminosity profile obeys the 
${\rm R^{1/4} }$
law. In such a case the relation between ${\rm R_{L}(t)}$ and  
${\rm M_{L}(t)}$ is

%%%%%%%%%%%%equation 9
\begin{equation}
{\rm R_{L}(t) = 26.1\times (M_L(t)/10^{12} M_{\odot})^{(2-\eta)} }
 ~~{\rm kpc}    
\label{lum_rag}
\end{equation}
with $\eta=1.45$ (cf. also Arimoto \& Yoshii 1987).

Finally, another useful quantity is the volume density of gas 
${\rm \rho_{g}(t)}$ given by 

%%%%%%%%%%%%%%equation 10
\begin{equation}
{\rm \rho_{g}(t) = {3 M_{g}(t) \over 4\pi R_{L}(t)^3}  }      
\label{rho}
\end{equation}

It is worth recalling that if dark and luminous matter are supposed to have the
same spatial distribution, equations (\ref{gas_pot}) and (\ref{dark_pot})
are no longer required. Indeed, the  binding energy of the gas can be simply
obtained from the total gravitational energy of Saito (1979b)
provided that the mass ${\rm M_{L}(t)}$ 
is replaced by the sum ${\rm M_{L}(t) + M_{D} }$.
In such a case, the binding energy of the gas is 
\verylittleskip 

%%%%%%%%%%%%%equation 11
\begin{equation}
{\rm \Omega_{g}(t) = \Omega_{L+D}(t) M_{g}(t) [2 - M_{g}(t)] }.  
\label{gas_new}
\end{equation}

\subsection{ Basic Equations of Chemical Evolution}\label{bas_equ} 

Denoting with ${\rm X_{i}(t)}$ the current mass abundance of an element $i$ and
introducing the dimensionless variables 

%%%%%%%%%%%%%%equation 12
\begin{equation}
{\rm G(t)=M_{g}(t)/M_{L}(T_{G})  }           
\label{gas_fra}
\end{equation}
\noindent
 and 

%%%%%%%%%%%%%%equation 13
\begin{equation}
{\rm G_{i}(t)=G(t)X_{i}(t),   }       
\label{gas_i}
\end{equation}
\noindent
the equations governing the time variation of the ${\rm G_{i}(t)}$ and hence 
${\rm X_{i}(t)}$ are
\verylittleskip

%%%%%%%%%%%%%%%equation 14
\begin{displaymath}
{\rm   {dG_{i}(t) \over dt}= - X_{i}(t) \Psi(t) ~~ +~~~~~~~~~~~~~~~~~~ }
\end{displaymath}
\begin{equation}
{\rm  \int_{M_L}^{M_U} \Psi(t-t_{M}) 
\phi(M) Q_{i}(t-t_{M}) dM + {d[G_{i}(t)]_{inf} \over dt }  }       
\label{degas_i}
\end{equation}
\noindent
where ${\rm \Psi(t)}$ 
is the rate of star formation in units of ${\rm M_L(T_G)}$, 
${\rm t_M}$ is
the lifetime of a star of mass ${\rm M}$, ${\rm \Phi(M)}$ is the initial mass 
function,
${\rm Q_{i}(t-t_{M})}$ is the fraction of mass ejected by such a star in 
form of an
element $i$, and ${\rm \dot{[G_{i}(t)]}_{inf} }$ is the rate of gas 
accretion. This is expressed by 

%%%%%%%%%%%%%%%%equation 15
\begin{equation}
{\rm {d[G_{i}(t)]_{inf} \over dt}= {X_{inf} \over M_g(t) } {dM(t) \over dt } }  
\label{gas_inf}
\end{equation}
 
By definition ${\rm \Sigma_i X_i(t)=1}$. All the details of the 
${\rm Q_{i}(t-t_{M})}$ are
omitted here for the sake of brevity. They can be found in Chiosi (1986) and
Matteucci (1991) to whom the reader should refer. 

The stellar birth rate, i.e. the number of stars of mass M born in the 
interval ${\rm dt}$ and mass interval ${\rm dM}$ is:

%%%%%%%%%%%%%%%equation 16
\begin{equation}
{\rm dN =\Psi(t,Z) \phi(M)~dM~dt }  
\label{birth}
\end{equation}
\noindent
where ${\rm \Psi(t)}$ is the rate of star formation as a function of time and
enrichment, while ${\rm \phi(M)}$ 
is the initial mass function (IMF). 

The IMF is the
Salpeter law expressed as 
\verylittleskip

%%%%%%%%%%%%%%%equation 17
\begin{equation}
{\rm \phi(M)=  M^{-x}  }  
\label{imf}
\end{equation}
\noindent
where ${\rm x=2.35}$. The IMF is
normalized by choosing the parameter $\zeta$
\verylittleskip

%%%%%%%%%%%%%%%%equation 18
\begin{equation}
{\rm \zeta =  \frac{\int_{M_*}^{M_U}\phi(M){\times}M{\times}dM}
{\int_{M_L}^{M_U}\phi(M){\times}M{\times}dM}   } 
\label{zeta}
\end{equation}
\noindent
i.e. the fraction of total mass in the IMF above ${\rm  M_{*}}$, and 
deriving the lower limit of integration ${\rm M_L}$. 
The upper limit of integration is ${\rm M_{U}=120 M_{\odot}}$, 
the maximum mass in
our data base of stellar models, while the mass limit  ${\rm M_*}$ 
is the minimum
mass contributing to the nucleosynthetic enrichment of the interstellar medium
over a time scale comparable to the total lifetime of a galaxy. This mass is
approximately equal to ${\rm 1 M_{\odot}}$. 

\subsection{ Star Formation Rate}\label{bas_sfr} 
The rate of star formation is assumed to depend on the gas mass according to 

%%%%%%%%%%%%%equation 19
\begin{equation}
{\rm \Psi(t)= \nu M_{g}(t)^{k} }      
\label{sfr}
\end{equation}
\noindent
where $\nu$  and $k$ are adjustable parameters. 
The star formation rate normalized to ${\rm M_L(T_G)}$ becomes

%%%%%%%%%%%%%equation 19/b
\begin{equation}
{\rm \Psi(t)= \nu M_{L}(T_G)^{k-1} G(t)^{k}  }     
\label{sfr1}
\end{equation}

Linear and quadratic dependencies of the star formation rate on the gas
content, $k=1$ and $k=2$ respectively, were first proposed by  Schmidt (1959)
and have been adopted ever since because of their simplicity (see Larson 1991
for a recent review). 

With the law of star formation of equation (\ref{sfr}), the time behaviour of
the rate of star formation ${\rm \Psi(t)}$  depends on the type of model in usage. In
the closed-box description, the rate of star formation is maximum at the
beginning, and since then  it continuously decreases  until galactic winds
occur. In the infall model, owing to the competition between the rate of gas
infall and gas consumption by star formation, the rate of star formation starts
small, increases to a maximum and then declines. The age at which the peak
occurs approximately corresponds to the infall time scale $\tau$. 
\littleskip

\subsection{ Galactic Winds }\label{sn_win} 

The models include the occurrence of galactic winds. When the energy deposit by
the supernova explosions and stellar winds from massive stars (see below)
exceeds the gravitational binding energy of the gas, the galactic wind is let
occur halting further accretion and star formation, and ejecting all remaining
gas. The mass of the luminous matter is  frozen to the current value of the
mass in stars. This is the real mass of the remnant visible galaxy to be used
to compare theoretical results with observations. However, for the sake of
simplicity we will always refer  to the models by means of the asymptotic mass
${\rm M_{L}(T_{G})}$. 

The galactic wind  occurs when the total thermal energy of the gas equates the
binding energy  of the gas. The thermal energy of the gas is determined by the 
energy input from  Type I and Type II supernovae and  stellar winds ejected by
massive stars (see also Theis et al. 1992). 

The  energy input from the three components are 
\verylittleskip

%%%%%%%%%%%%equation 20
\begin{equation}
{\rm E_{th}(t)_{SNI} =  \int_{0}^{t} \epsilon_{SN}(t-t') R_{SNI}(t') 
              M_{L}(T_G) dt'  }    
\label{esnI}
\end{equation}

%%%%%%%%%%%%equation 21
\begin{equation}
{\rm E_{th}(t)_{SNII} =  \int_{0}^{t} \epsilon_{SN}(t-t') R_{SNII}(t')
              M_{L}(T_G) dt' }    
\label{esnII}
\end{equation}

%%%%%%%%%%%%%equation 22
\begin{equation}
{\rm E_{th}(t)_{W} =  \int_{0}^{t} \epsilon_{W}(t-t') R_{W}(t')
              M_{L}(T_G) dt' }    
\label{eW}
\end{equation}
\noindent
where ${\rm R_{SNI}}$ and ${\rm R_{SNII}}$ are the rates of supernova 
production, while
${\rm R_{W}}$ is the rate of gas ejection by massive stars in form of 
stellar winds. 
The time ${\rm t'}$ is either the explosion time or the time at which 
stellar winds occur. 

To calculate ${\rm \epsilon_{SN}(t)}$ and 
${\rm \epsilon_{W}(t)}$, 
the thermal content in the supernova remnant and stellar 
wind, respectively, we proceed as follows.
The time variation of the supernova thermal content is  
taken  from  Cox (1972) 

%%%%%%%%%%%%%equation 23
\begin{equation}
{\rm \epsilon_{SN}(t_{SN}) = 7.2 \times 10^{50} \epsilon_0 }    ~~{\rm erg}
\label{eps_sn}
\end{equation}
\noindent  
for ${\rm 0 \leq t_{SN} \leq t_c }$  and
 
%%%%%%%%%%%%%%equation 24
\begin{equation}
{\rm \epsilon_{SN}(t_{SN}) = 2.2 \times 10^{50} 
               \epsilon_{0}(t_{SN}/t_{c})^{-0.62}  }     ~~ {\rm erg} 
\label{t_sn}
\end{equation}
\noindent
for ${\rm t_{SN} \geq t_c }$, where $\epsilon_0$ is the initial blast wave 
energy of
the supernova in units of $10^{51}$ ergs, which is assumed equal to unity for
all supernova types, ${\rm t_{SN}=t-t'}$ is the time elapsed since the 
supernova
explosion, and ${\rm t_c}$  is the cooling time of a supernova remnant, i.e. 
\verylittleskip

%%%%%%%%%%%%equation (25)
\begin{equation}
{\rm t_{c}=5.7 \times 10^4 \epsilon_{0} ^{4/17} n_{0}^{-9/17} } ~~{\rm yr}, 
\label{tcool_sn}
\end{equation}
\verylittleskip
\noindent
where ${\rm n_{0}(t)=\rho_{g}(t)/m_H }$ is the number density of the 
interstellar
medium as a function of time, and ${\rm \rho_{g}(t)}$ is the corresponding gas
density. 

The thermal content in the wind ejected by a massive star is estimated in the
following way: a typical massive star (say in the range  ${\rm 20 M_{\odot}}$ 
to ${\rm 120 M_{\odot}}$) in the course of its evolution ejects about 
half of the  mass in
form of a wind with terminal velocity of about $2000~ {\rm Km ~s^{-1}}$ (see
Chiosi \& Maeder 1986) and, therefore, injects into the interstellar medium an
energy of about 
\verylittleskip

%%%%%%%%%%%%equation (26)
\begin{equation}
{\rm  \epsilon_{W0} = {1 \over 2} ({M \over 2})V^2({Z \over Z_{*}})^{0.75} }   
                      ~~{\rm erg},   
\label{e_wo}
\end{equation}
\verylittleskip
\noindent
where the term ${\rm Z/Z_{*}}$ takes into account a possible dependence 
on the metal
content (cf. Kudritzki et al. 1987, Theis et al. 1992, and BCF for details). 

Part of this energy will be radiated away and part will contribute to heat up
the gas. In analogy with  supernova remnants, we assume that this energy will
cool down with the same law as for the energy injected by supernovae but with a
different cooling time scale $t_{cw}$. This assumption stands on the notion
that the energy content in the wind is continuously refueled by the radiative
energy output from the emitting star. Therefore 

%%%%%%%%%%%%%%equation  27
\begin{equation}
{\rm \epsilon_{W}(t_{W}) = \epsilon_{W0} }    ~~{\rm erg}  ~~~
\label{eps_w}
\end{equation}
\noindent
for ${\rm  0 \leq t_{W} \leq t_{cw} }$ and 

%%%%%%%%%%%%%%equation  28 
\begin{equation}
{\rm \epsilon_{W}(t_{W}) =  \epsilon_{W0}(t_{W}/t_{cw})^{-0.62}  }
     ~~ {\rm ergs} 
\label{eps_w_cool}
\end{equation}
\verylittleskip
\noindent
for ${\rm t_{W} \geq t_{cw}}$, where ${\rm t_{W}=t-t'}$ 
is the time elapsed since the birth
of a massive star, and ${\rm t_{cw}}$ is supposed to be of the same order  
of the
evolutionary time scale of a massive star. Many preliminary calculations show
that ${\rm t_{cw}=15\times 10^6}$ yr is a good choice. It is easy to figure 
out that
stellar winds can contribute by as much energy as classical supernova
explosions and, therefore, they cannot be neglected in evaluating the total
energy required to power galactic winds. 

Other details on the calculation of ${\rm E_{th}(t)_{SNI}}$, 
${\rm E_{th}(t)_{SNII} }$ 
and ${\rm E_{th}(t)_{W} }$, the frequency of type I SN in particular can be 
found  in BCF to whom we refer.

Suffice to remark that owing to the different metallicity distribution of 
infall models with respect to the closed box ones --- it is indeed skewed 
towards the high metallicity end --- the metallicity dependence of 
eq.(\ref{e_wo}) can be neglected. For the sake of consistency also the 
closed-box models presented here are calculated with same assumption

The total thermal energy stored into the interstellar medium is 

%%%%%%%%%%%%equation 29
\begin{equation}
{\rm E_{th}(t) = E_{th}(t)_{SNI} + E_{th}(t)_{SNII} + E_{th}(t)_{W}.  } 
\label{en_tot}
\end{equation}
 
Finally, a galactic wind is supposed to occur when 

%%%%%%%%%%%%%equation 30
\begin{equation}
{\rm E_{th}(t) \geq \Omega_{g}(t).  }  
\label{wind}
\end{equation}

This is an oversimplified description of the effect of the heating processes on
triggering galactic winds. It implicitly assumes that the energy deposit is
immediately homogenized in the gas, so that condition (\ref{wind}) is met at a
certain age by  the gas as a whole.  In reality, the situation is much more
complicated because cavities and tunnels of hot gas are likely to  form
(Matthews \& Baker 1971),  so that condition (\ref{wind}) can be met only
locally and fractions of gas are ejected. This would imply that galactic winds
rather than instantaneously take place over certain time scales that are
difficult to evaluate at least in a simple model like this. 

Furthermore, the deposit of thermal energy by supernova explosions and stellar
winds, may also directly affect star formation. Indeed it is conceivable that
as gas gets hotter and hotter, star formation gets less efficient as well. 

Despite the above comments, all  the results below are derived from models in
which gas heating  has no direct effect on  star formation and
galactic winds occur instantaneously when condition (\ref{wind}) is met by the
gas as a whole. 

Finally, the gas  continuously 
 ejected by dying stars is supposed to quickly heat up at high
temperatures corresponding to the high velocity dispersion of the galaxy 
and not to able to form new stars
The onset of
galactic winds halts any star formation activity. 

\subsection{ Model Parameters}\label{par_mod}  

Primary parameters of the models are:

1) The galactic mass ${\rm M_{L}(T_{G})}$. This is a mere label identifying the
models. In the closed-box models it is the initial luminous mass of the galaxy,
whereas in the infall models it represents the asymptotic value. In any case 
galactic winds lower the galactic mass to the value already stored in stars at
the age of wind ejection. Models (closed-box and infall) are calculated for the
following values of   ${\rm M_{L}(T_{G})}$:  $3\times10^{12}$\Msun,
$1\times10^{12}$\Msun, $5\times10^{11}$\Msun, $1\times10^{11}$\Msun,
$5\times10^{10}$\Msun, and $1\times10^{10}$ \Msun. Thereinafter, all the models
will be labelled by the mass ${\rm M_{L}(T_{G})}$ in units 
of $10^{12}$ \Msun\ in
turn shortly indicated by \M12. 

2) The ratios ${\rm R_{L}/R_{D}}$ and ${\rm M_{L}/M_{D}}$ which 
fix the gravitational potential
and the effect of dark matter. They have  already been described in
\S~\ref{out_mod} 

3) The exponent $k$ and efficiency $\nu$ of the star formation rate. All the
models are calculated assuming for $k=1$. The parameter $\nu$ and associated
time scale of stars formation ${\rm t_{SF}=1/\nu}$ are varied in order 
to fit the major constraints of the problem. 

4) The initial mass function (slope and $\zeta$). The slope is kept constant,
whereas $\zeta$ is let vary. We will show that $\zeta=0.5$ is a good choice in
order to get models with  ${\rm M/L_{B}}$ ratios in agreement with the 
observational
data (see below). This value is compatible with current prescriptions for
chemical yields, indicating the range $0.3 < \zeta < 0.50$ (cf. Matteucci 1991,
 Rana 1991, and Trimble 1991). 

5) The infall time scale $\tau$. Although the time scale of mass accretion is
considered as a  free parameter of the models, we guess its value from the
collapse time scale as function of ${\rm M_L(T_G)}$. Assuming a homogenous
distribution of matter in the volumes filled by the dark and luminous
components, we can express the  total mass density ${\rm \rho_T}$ as 

%%%%%%%%%%equation 31
\begin{equation}
{\rm \rho_T= (1 + \beta^2) \rho_L  }
\end{equation}
where ${\rm \rho_L}$ can be determined with the aid of equation (10). 
Accordingly,
 the collapse time scale  is (see also Arimoto \& Yoshii 1987) 

%%%%%%%%%%equation 32 %%%%%%%%%%%%
\begin{displaymath}
{\rm \tau(M_L(T_G)) = (1 + \beta^2)^{-0.5}  } 
\end{displaymath}
\begin{equation}
~~~~~~~~~~~~~\times 
     {\rm   1.26 \times 10^8 ({ M_L(T_G)\over {10^{12}M_{\odot}}})^{0.325} }  
                       ~~{\rm yr}
\end{equation}
\noindent
This relation yields  the collapse time scales  reported in Table~3 as function
of the galactic mass. The time scale of mass accretion rules the infall
process. For $\tau$ going  to 0, the model reduces to an initial burst of star
formation, while for $\tau$ stretching to $\infty$, star formation does never
stop. However, all the models below are calculated with $\tau=0.1$ Gyr
independently of the galactic mass. This choice reduces the number of
parameters and appears fully appropriate to the present purposes. 

6) The galaxy age ${\rm T_{G}}$. This  could be constrained to  the model of the
universe, i.e. to  ${\rm H_{0}}$, ${\rm q_{0}}$, and red-shift of galaxy 
formation
${\rm z_{for}}$. However, 
because of the uncertainty on  the cosmological parameters,
 galaxy  ages are let vary in the range of current estimates
of the globular cluster ages, i.e. $13 \div 15 \pm 2$ Gyr
  (cf. Chiosi et al. 1992 for a recent review of the
subject). 

\setcounter{table}{2}
%%%%%%%%%Table 3(il tempo scala di free-fall) %%%%%
\begin{table}[htb]
\vskip 0.3 cm
%%%%%%%%%%%%%%%%%%%%%%%%%%%%%%%%%%%%%%%%%
\begin{center}
\vskip 0.2 cm
\caption{{\em Free-Fall} time scale for the various masses }
\vskip 0.25 cm
\scriptsize
\begin{tabular} {c| c c c c c c}
\hline
\hline
 & & & & & & \\
\M12 & 3     & 1    & 0.5  & 0.1  & 0.05 & 0.01 \\
 & & & & & & \\
\hline
 & & & & & & \\
${\rm \tau_{ff}}$ (Gyr) & 0.18 & 0.12 & 0.10 & 0.06 & 0.05 & 0.03 \\
 & & & & & & \\
\hline
\hline
\end{tabular}
\end{center}
\label{tab3}
\end{table}

\subsection{General remarks}

Before concluding this section, we like to call the attention on three
drawbacks of the models that might be used to invalidate the present analysis.
The points in question are (i) the adoption of the one-zone description which
does not allow us to take into account spatial gradients in colours and
metallicity, (ii) the solar partition of abundances used in the stellar models,
 and finally (iii) our  treatment of galactic winds. 

(i) It is clear that the  \VK\ colours and magnitudes of the galaxies in the
Bower et al. (1992a,b) sample  refers to the whole galaxy, whereas the UV
excess and its companion colour \UVex\ of Burstein et al. (1988) in most cases
refer to the central region of a galaxy. The existence of gradients in colours
and metallicities across elliptical galaxies is a well established
observational fact (Carollo et al. 1993; Carollo \& Danziger 1994, Davies et
al. 1993, Schombert et al. 1993). This means that insisting on the one-zone
description as in the present models, could be source of difficulty when
comparing model results with observational data. It seems reasonable to argue
that higher metallicities can be present in the central regions thus
facilitating  the formation of the right type of stellar sources of the UV
radiation (H-HB and AGB manqu\'e stars), whereas lower metallicities across the
remaining parts of the galaxies would make the other integrated colours such as
(V--K) bluer than expected from straight use of the one-zone model for the
whole galaxy. 

(ii) Recent observations indicate that the pattern of abundances in elliptical
galaxies is skewed towards an overabundance of $\alpha$-elements with respect
to Fe (cf. Carollo et al. 1993; Carollo \& Danziger 1994). As already recalled,
 the library of stellar models in use is based on the standard (solar) pattern
of abundances. 
Work is in progress to generate libraries of stellar models and isochrones with 
 [${\rm \alpha/Fe] >0 }$. Preliminary calculations for the solar metallicity 
(Z=0.02) show the effect to be small in the theoretical CMD. 
In addition, calculations of the ${\rm Mg_2}$ index from synthetic spectra for 
different metallicities and partitions of $\alpha$-elements (Barbuy 1994) show 
that passing from ${\rm [\alpha /Fe]=0.0 }$ to ${\rm [\alpha /Fe]=0.3 }$ 
the ${\rm Mg_2}$ index of an old SSP (15 Gyr) increases from 0.24 to 0.30.
\littleskip

(iii) Concerning galactic winds, we would like to  comment   on the recent
claim by Gibson (1994) that using the above formulation for the energy
deposited by supernova explosions and stellar wind, BCF underestimate the
effect of supernova explosions and on the contrary overestimate the effect of
stellar winds from massive stars. In brief,  BCF assuming  the CMR to be a
mass-metallicity sequence looked at the metallicity
that would generate the right colours.  They found that considering supernovae
as the only source of energy, by the time galactic winds occur,  the gas
fraction has become too low and the metallicity too high in turn, that the CMR 
is destroyed (it runs flat). To cope with this difficulty they included another
source of energy, i.e. the stellar winds from massive stars whose 
efficiency in powering galactic winds is described by equations 
(25) through (27)
with  ${\rm t_{cw}}$ as the key parameter for which the value of  
 $1.5\times 10^7$ yr has been adopted. 
This time roughly corresponds to the evolutionary lifetime of a 10 
${\rm M_{\odot} }$ star. In other words it is the time scale over which a 
group of newly formed O-type stars would evolve away from the SSP. They found
that ${\rm t_{cw}}$ shorter than this value  would not allow  sufficient
powering of the interstellar medium, so that galactic winds  would occur much
later than required by the CMR. Finally, BCF provided also an explanation for
the different results found by other authors (Arimoto \& Yoshii 1987, Matteucci
\& Tornamb\'e 1987, Angeletti \& Giannone 1990, and Padovani \& Matteucci 1993).
 Admittedly, this additional source of energy was invoked to keep the standard
interpretation of the CMR.

\section{Closed-Box  models of elliptical galaxies}\label{clo_mod}

In order to test the dependence of the results on the new library of SSPs we
have re-calculated models of elliptical galaxies with the closed-box
approximation. The guide line of the analysis is to search the combination of
the parameters $\nu$ and $\zeta$ for which the most massive
galactic model (3\M12) at the canonical age of
15 Gyr  matches the maximum observational values for 
(V--K) and (1550--V) colours, 3.5 and 2 respectively (cf. Figs.\ref{fig13}
and \ref{fig16} below).  To this aim, assuming
$k=1$, different values of $\nu$ and $\zeta$ are explored. 

\begin{figure}
%\picplace{ 9cm}
\psfig{file=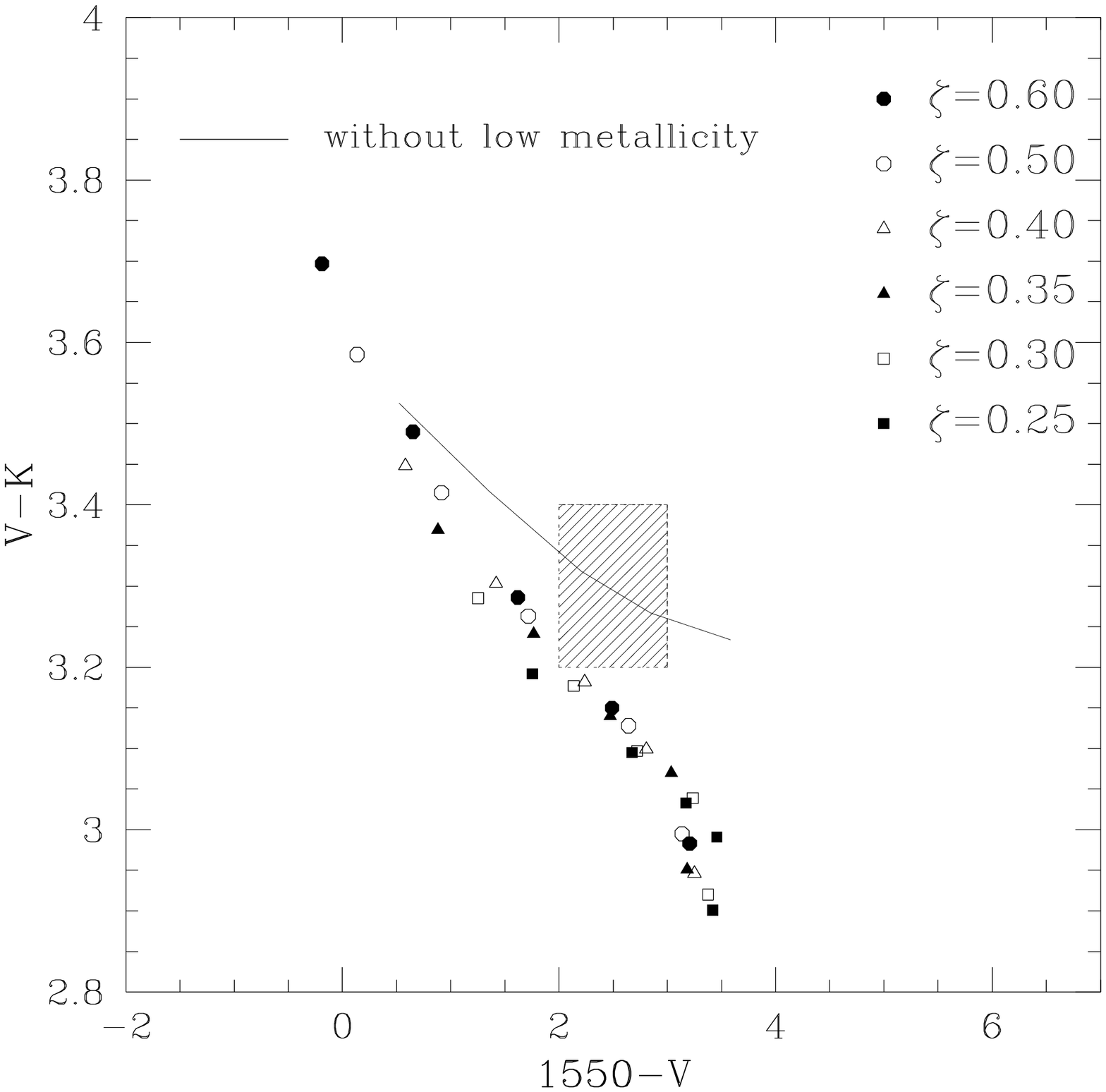,height=9.0truecm,width=8.5truecm}
\caption{The (1550--V) versus (V--K) relationship  for closed-box  models. The
shaded area shows the observational data. The models are calculated for
different values of $\zeta$ as indicated, and for all the values of $\nu$
listed in Table~4.  The solid line shows  models with  $\zeta=0.40$ and the
same values of $\nu$ as before but in which the low-metallicity component is
taken away. All the models refer to the 3\M12 galaxy } 
\label{fig5}
\end{figure}

In the closed-box approximation, $\nu$ simply fixes the time scale of star
formation. The values under consideration are listed  in Table~4. Furthermore,
for each value of $\nu$, six values of $\zeta$ are taken into account, i.e.
$\zeta=0.25$, 0.30, 0.35, 0.40, 0.50, and 0.60. The analysis is limited to the
case of the 3\M12\ galaxy. The details of these models are not given here for
the sake of brevity; they are  available from the authors upon request. The
results are shown in Fig.~\ref{fig5} which displays the correlation between
(V--K) and (1550--V) and the corresponding observational data (shaded area).
The (1550--V) colours are from Burstein et al.  (1988), whereas the (V--K)
colours are from Bower et al. (1992a,b).

%%%%%%%%%Table 4 (il parametro nu e rispettivi tempiscala di SFR) %%%%%
\begin{table}[htb]
\vskip 0.3 cm
%%%%%%%%%%%%%%%%%%%%%%%%%%%%%%%%%%%%%%%%%
\begin{center}
\vskip 0.2 cm
\caption{Adopted values for the  parameter $\nu$ and associated
 time scale of star formation ${\rm t_{SF}}$  }
\vskip 0.25 cm
\scriptsize
\begin{tabular} {c| c c c c c}
\hline
\hline
 & & & & & \\
$\nu$ (${\rm Gyr^{-1}}$)& 20    & 15    & 10    & 5     & 1    \\
 & & & & & \\
\hline
 & & & & & \\
${\rm t_{SF}}$ (${\rm Gyr}$) & 0.050 & 0.067 & 0.100 & 0.200 & 1.000 \\
 & & & & & \\
\hline
\hline
\end{tabular}
\end{center}
\label{tab4}
\end{table}

Clearly no combination of the parameters is found yielding models 
satisfying the above constraint (upper left corner of the shaded area in 
Fig. \ref{fig5}).  
 Varying the age within the range allowed by the CMR 
(say $15 \pm 2$ Gyr) does not help to match both colours of the reference
galactic model. Similar results are obtained for the other
galactic masses.
There is a tight relationship between the theoretical
(1550--V) and (V--K) which can be easily understood in terms of intrinsic
behaviour of the closed-box model and the dependence of the two colours on the
metallicity. As shown by BCF, the maximum metallicity determines the UV
intensity and hence (1550--V) colour, whereas the mean metallicity governs the
(V--K) colour. 

%%%%%%%%Figure 5 %%%%%%%%%%%

In a closed-box system, the current (maximum)  metallicity obeys the relation: 

%%%%%%%%%%%%equation 33
\begin{equation}
{\rm   Z(t) = Y_{z} \ln{\mu^{-1}} }
\label{zy}
\end{equation}
where ${\rm \mu = M_{g}(t)/M_{L}(T_G)}$ 
is the fractionary mass gas and ${\rm Y_{z} }$ is the
chemical yield per stellar generation. This relation holds for ${\rm Z \ll 1}$. 
The mean metallicity is  given by:

%%%%%%%%%%%%equation 34
\begin{equation}
{\rm   \langle Z(t) \rangle = {{\int \Psi(t) Z(t) dt} \over {\int \Psi(t) 
                     dt}}   } 
\label{zmed}
\end{equation}

\begin{figure}
%\picplace{ 9cm}
\psfig{file=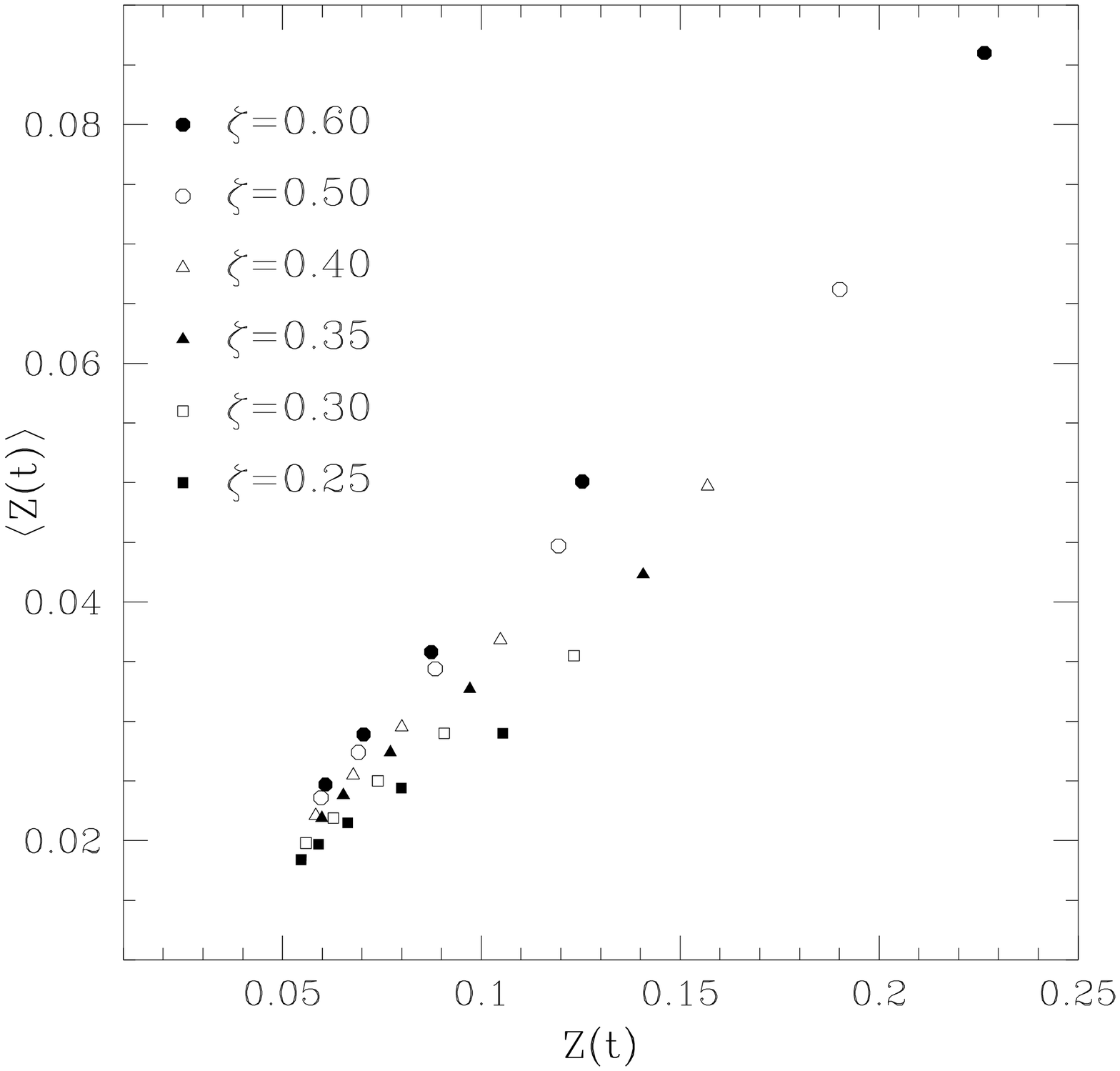,height=9.0truecm,width=8.5truecm}
\caption{ The ${\rm Z(t)}$ versus ${\rm \langle Z(t) \rangle}$ 
for the same models as in
Fig.~\protect\ref{fig5}. Note the linear relationship between the two
metallicities for models with the same $\zeta$. See the text for more details} 
\label{fig6}
\end{figure}

With the aid of eqs. (\ref{zy}), (\ref{zmed}) and (\ref{sfr}) we get the
relationship between ${\rm \langle Z(t) \rangle }$ and ${\rm Z(t)}$ shown in
Fig.~\ref{fig6}. Owing to the low-metal skewed distribution typical of the
closed-box models, if the maximum metallicity rises to values at which the H-HB
 and AGB manqu\'e stars are formed and in turn the dominant source of the UV
radiation is switched on (cf. BCF for all details), the mean metallicity is to
low to generate the right (V--K) colours. In order to prove this statement, in
Fig.~\ref{fig5} we show models in which all the  stars with $Z \leq 0.008$ are
artificially removed (solid line). As expected, solutions are now possible. On
the one hand this result confirms the conclusions by BCF about the intrinsic
difficulty of the closed-box models, on the other hand it makes evident that
the success or failure of a particular spectro-photometric model in matching
the observational data depends very much on details of the adopted spectral
library. With the one used by BCF reasonable match of the data was possible
even with the closed-box scheme, but for the before mentioned marginal
discrepancy of the ISED in the region 2000 \AA\ to 3500 \AA. With the new
library this is no longer feasible. 

%%%%%%%%Figure 6 %%%%%%%%%%%

\section{Infall models of elliptical galaxies }\label{result}

\begin{figure}
%\picplace{ 9cm}
\psfig{file=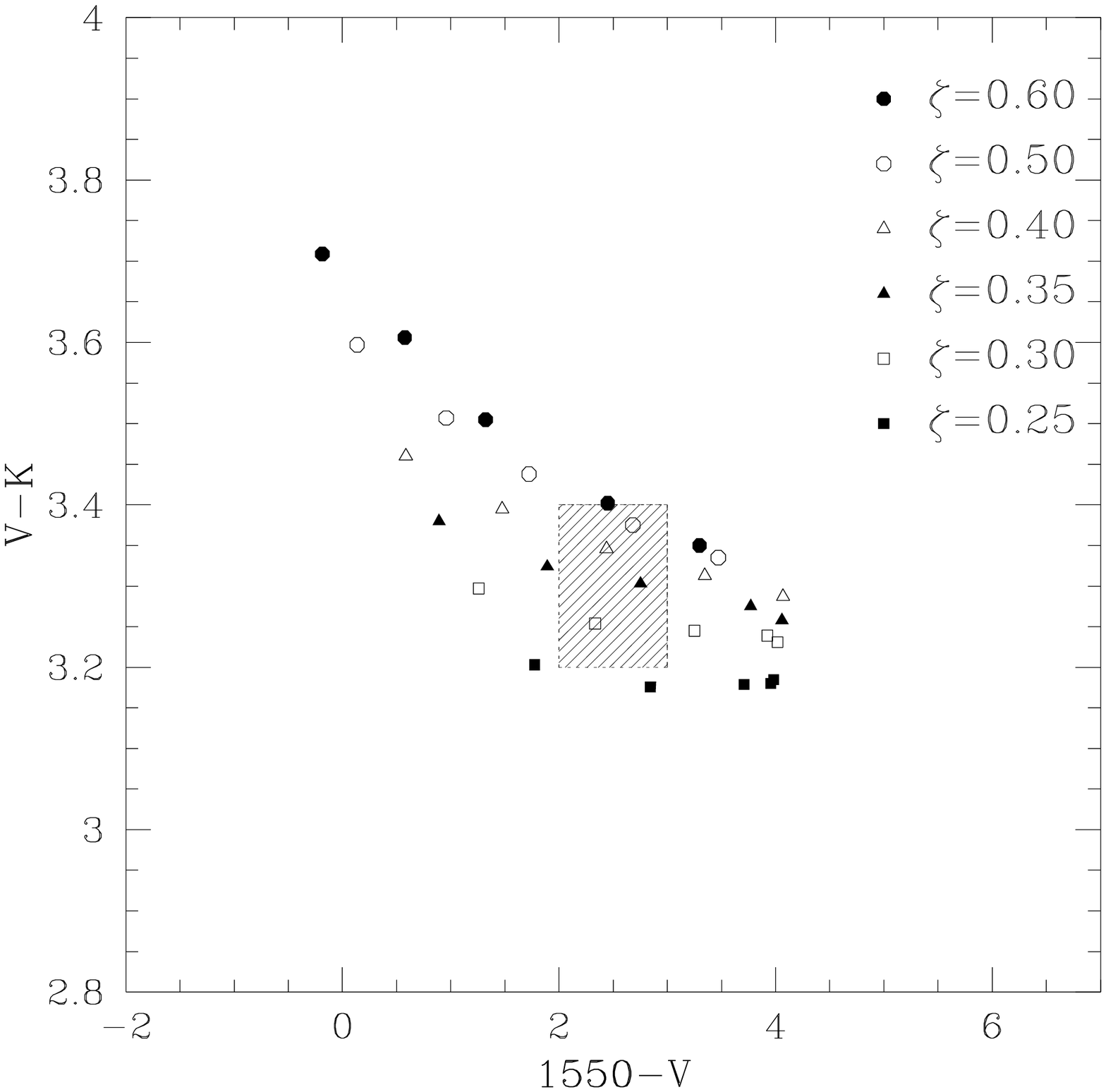,height=9.0truecm,width=8.5truecm}
\caption{The (1550--V) versus (V--K) for all infall models. The shaded area
indicates the range spanned by the observational data.  The models are
calculated for different values of $\zeta$ as indicated, and for all the values
of $\nu$ listed in Table~4.   All the models refer to the 3\M12 galaxy  } 
\label{fig7}
\end{figure}

In order to check whether infall models can cope with the above difficulty, we
perform a preliminary analysis of the problem calculating infall models with
the same parameters $\nu$ and $\zeta$ as for the closed-box case. In these
models $k=1$, $\tau=0.1$ Gyr, and the galactic mass is 3\M12. The results are
shown in Fig.~\ref{fig7}. With the infall scheme many models exist whose (V--K)
and (1550--V) match the observational data. 

%%%%%%%%Figure 7 %%%%%%%%%%%

%%%%%%%%Figure 8 %%%%%%%%%%%
\begin{figure}
%\picplace{ 9cm}
\psfig{file=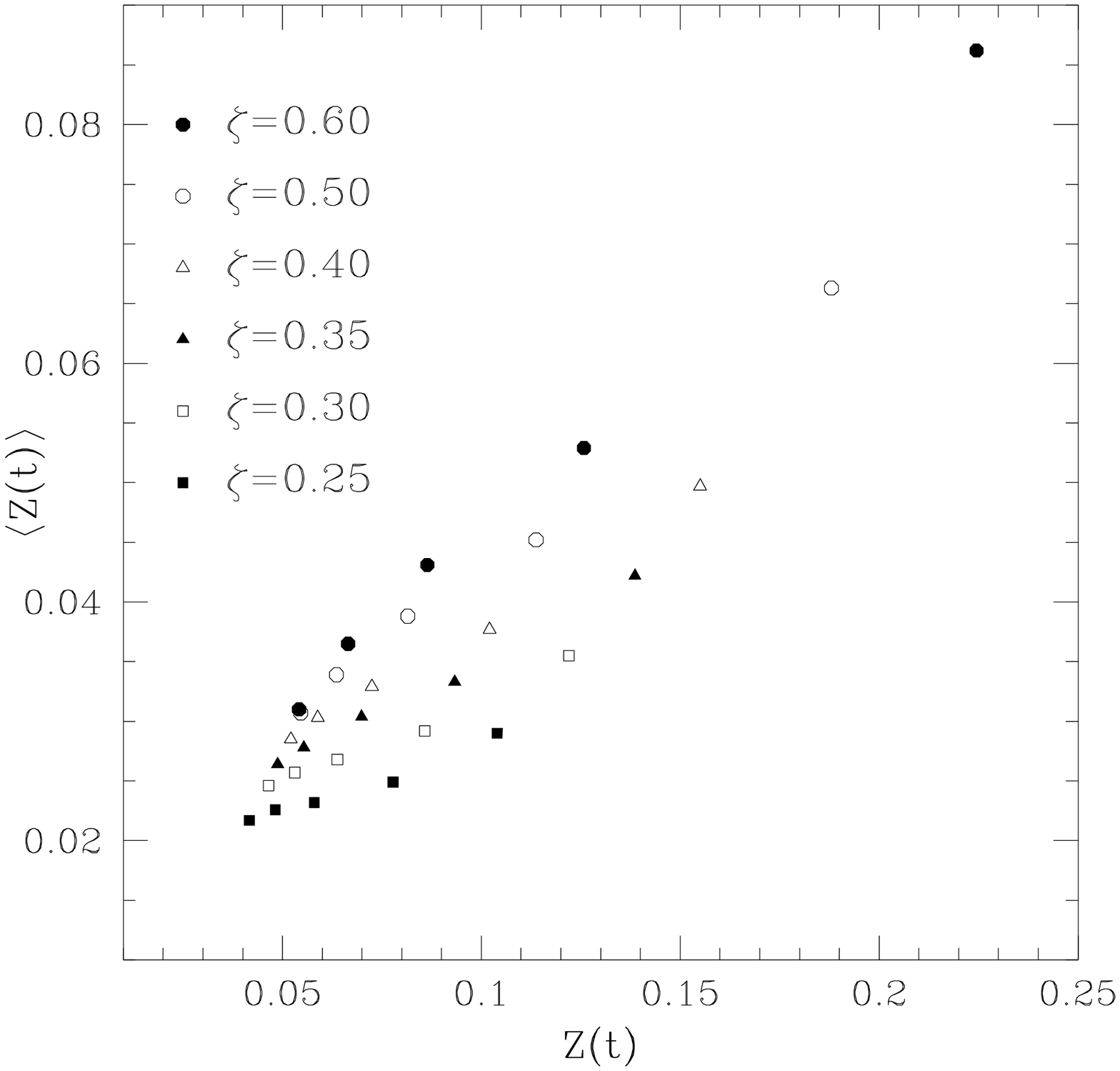,height=9.0truecm,width=8.5truecm}
\caption{ The ${\rm Z(t)}$ against ${\rm \langle Z(t) \rangle}$ 
for the open models as in
Fig.~\protect\ref{fig7}. Note the linear relationship between the two
metallicities for models with the same $\zeta$. See the text for more details} 
\label{fig8}
\end{figure}

The reason of the success is due to the law of chemical enrichment of 
infall models (cf. Tinsley 1980b)

%%%%%%%%%%%%%equation 35
\begin{equation}
{\rm  Z(t) = Y_{z} [ \ln (\frac{1}{\mu}) - 1]  }  
\label{newzy}
\end{equation}
where all the symbols have the same meaning as in eq. (\ref{zy}), which leads 
to the relationship
between the maximum and mean metallicity  shown in Fig.~\ref{fig8}. Since the
metallicity distribution of infall models is skewed towards the high
metallicity side, at given maximum metallicity higher mean metallicities are
found that allow us to match of the observational (V--K) and (1550--V) colours
at the same time. The details of these models are not given here for the sake
of brevity. They are available upon request. 

\section{\bf Properties of infall models }

Not all the models presented in the previous section are equally acceptable
because other important constraints of the problem have not yet been taken into
account. These are the slope and mean colours of the 
CMR for elliptical galaxies, the  mass to blue
luminosity ratio ${\rm M/L_{B}}$,  the overall fit of the UV excess, i.e. colour
(1550--V), shape of the ISED, and ${\rm Mg_{2}}$ versus (1550--V) relation. 

It turns out that  the ratio ${\rm M/L_{B}}$ requires $\zeta=0.5$, whereas the
combined effect of all the remaining constraints confines $\nu$ in the range $1
< \nu < 12$. More precisely, from $\nu=12$ for the 3 \M12  galaxy to $\nu=1$
for the  0.01 \M12 object. The various points will be examined in more
detail below. 

The main properties of the models in question are given in Table 5. This
consists of two parts. The first one contains the key quantities characterizing
the models at the stage of wind ejection: columns (1), (2) and (3) are the the
parameters $\nu$, $\zeta$, and ${\rm M_{L}(T_{G})}$ identifying the model,
respectively; column (4) is the age in Gyr at which the galactic wind occurs;
columns (5) and (6) are the fractionary mass of gas ${\rm G(t)}$ 
and living stars
${\rm S(t)}$ both in units of ${\rm M_{L}(T_{G})}$;  columns (7) and (8) 
are the maximum
and mean metallicity, respectively; column (9) is the rate of  star formation
${\rm \Psi(t)}$ (in ${\rm M_{\odot}/yr }$); 
finally columns (10) through (14) are the
binding energy of the gas (${\rm \Omega_{g}}$), 
the total thermal energy of the gas
(${\rm E_{g}}$), and the contribution to it by type I supernovae, 
type II supernovae,
and stellar winds, respectively. All energies are in units of $10^{50}$ ergs.
The second part of Table 5 presents the integrated magnitudes and colours of
the model galaxies at five different ages. Columns (1), (2) and (3) identify
the model by means of $\nu$, $\zeta$, and ${\rm M_{L}(T_{G})}$, 
respectively; column
(4) is the age in Gyr, columns (5) and (6) are the integrated absolute
bolometric (${\rm M_{bol}}$) and visual magnitude (${\rm M_{V}}$); 
finally, columns (7)
through (11) are the integrated colours (U--B), (B--V), (V--R), (V--K) and
(1550--V), respectively. We remind the reader that all the integrated
magnitudes refer to the current amount of the galaxy mass in form of living
stars (see below for more details). 

\subsection{Chemical Structure}
\begin{figure}
%\picplace{ 9cm}
\psfig{file=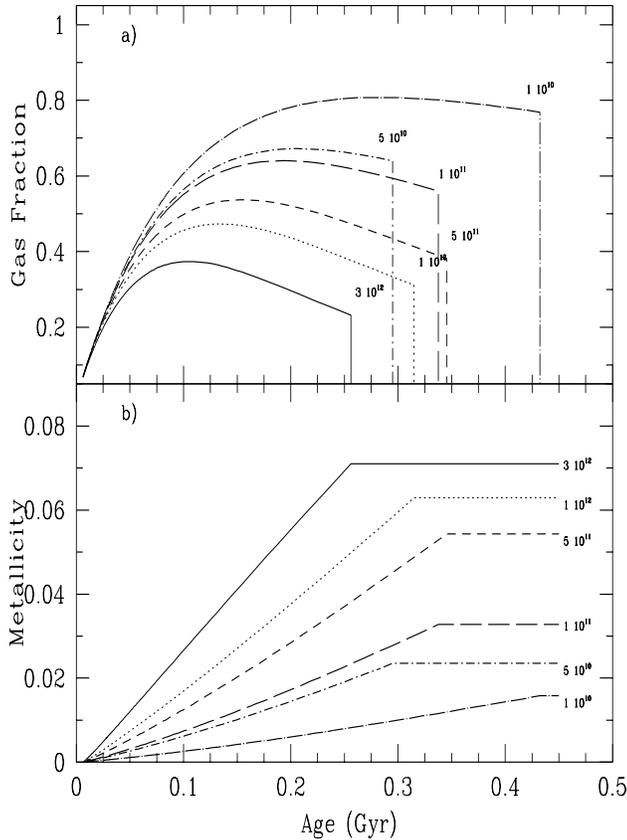,height=12.0truecm,width=8.5truecm}
\caption{ The gas fraction ${\rm G(t)}$ (Panel a) and metallicity 
${\rm Z(t)}$ (Panel b) as a function of the age in Gyr 
for the model galaxies with $\zeta=0.50$, 
$\nu$ increasing with the mass, and  galactic mass from  3 \M12\ to 
0.01 \M12.
In Panel (a), the intersection of the curves with the vertical lines
indicates the stage of the onset of galactic winds and consequent drop-off of
the gas content. In Panel (b) the horizontal lines show the constant
metallicity after the interruption of star formation by galactic winds   } 
\label{fig9}
\end{figure}

\begin{figure}
%\picplace{ 12cm}
\psfig{file=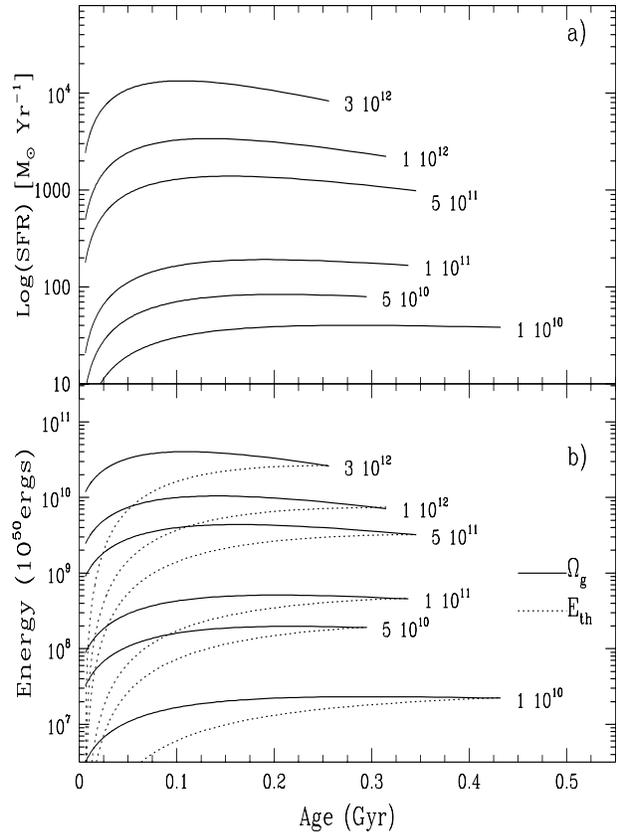,height=12.0truecm,width=8.5truecm}
\caption{ The same as in Fig.~\protect\ref{fig9} but for star formation rate
${\rm \Psi(t)}$ in units of \Msun yr$^{-1}$ (Panel (a)) and the gravitational
binding energy ${\rm \Omega_{g}(t)}$ and the thermal content 
of the gas ${\rm E_{th}(t)}$,
both in units of ${\rm 10^{50}~erg }$ (Panel (b))     } 
\label{fig10}
\end{figure}

 The time variation of the fractionary gas content $G(t)$ and metallicity
${\rm Z(t)}$ are shown in Fig.~\ref{fig9}, whereas that of the  star 
formation rate
(in units of ${\rm M_{\odot}/yr }$), thermal ${\rm (E_{th})}$ and 
binding energy of the gas
${\rm \Omega_{g}(t)}$ (both in units of $10^{50}$ erg) are displayed in
Fig.~\ref{fig10}. 

As expected, the rate of star formation (Panel a of Fig.~\ref{fig10}) starts
very small, grows to a maximum, and were it not  for the onset of galactic wind
and consequent interruption of the stellar activity, it would decline again.
The initial period of very low activity is the reason why infall models avoid
the so called G-Dwarf problem. 

\begin{figure}
%\picplace{ 9cm}
\psfig{file=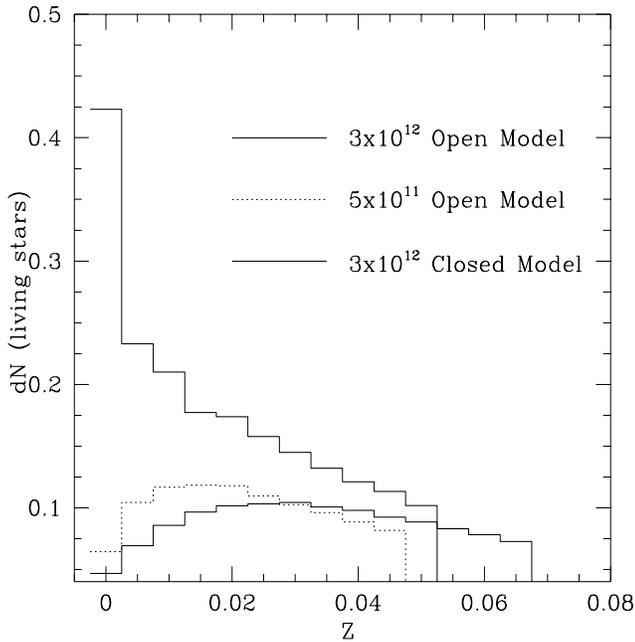,height=9.0truecm,width=8.5truecm}
\caption{ The relative number of alive stars per metallicity bin normalized to
${\rm M_{L}(T_{G})}$. 
The solid and dotted lines refer to infall models with 3\M12\ and $\nu =
12$, and  0.5 \M12\ and $\nu = 5.2$, respectively. Both are calculated with
$\zeta=0.50$. The thick line displays the closed-box model with 3\M12, 
$\nu=20$, and $\zeta = 0.40$} 
\label{fig11}
\end{figure}

\begin{figure}
%\picplace{ 9cm}
\psfig{file=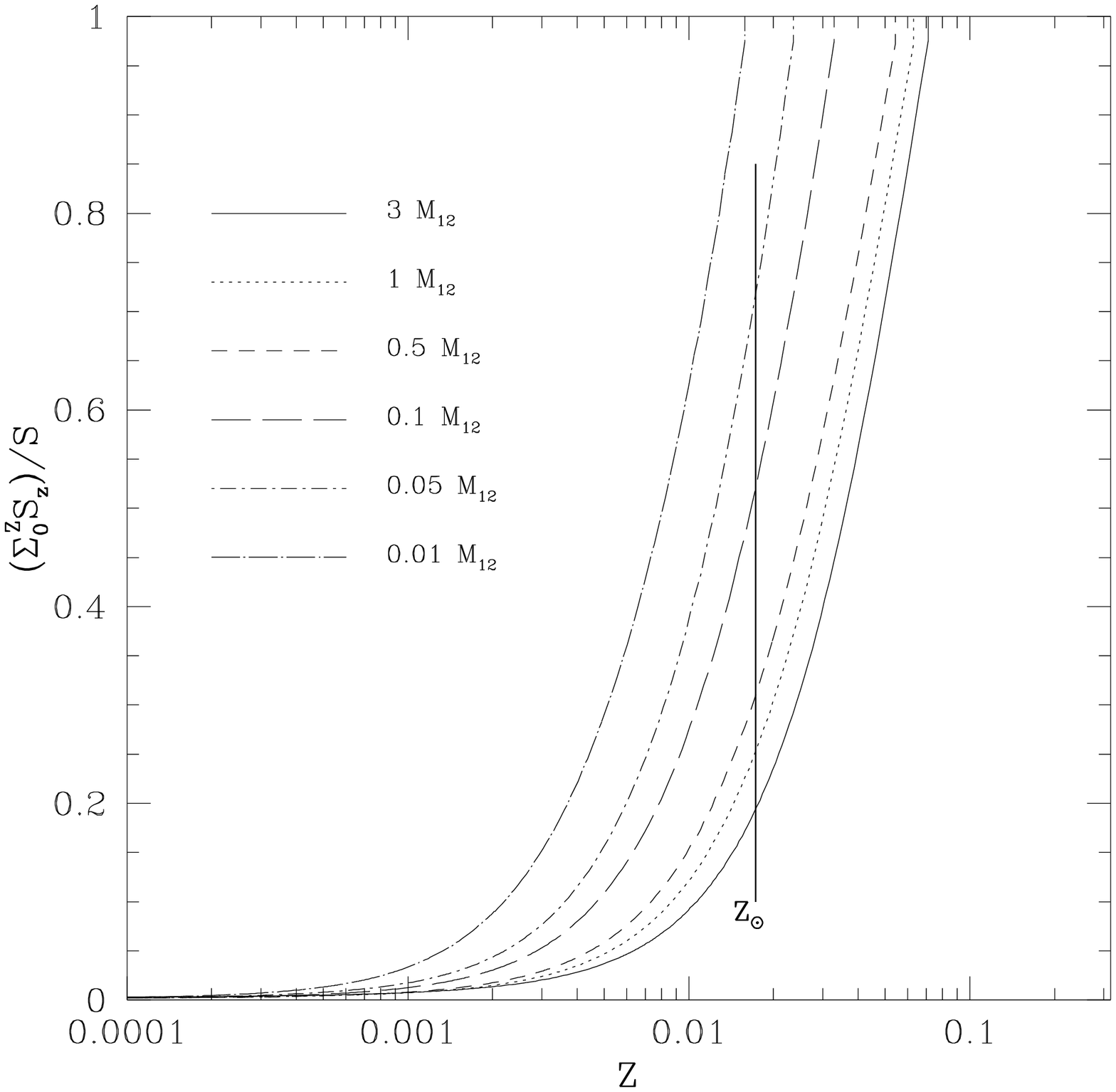,height=9.0truecm,width=8.5truecm}
\caption{ The cumulative fractionary mass of alive stars as function of the
metallicity for the model galaxies with  mass equal to 3, 1, 0.5, 0.1, 0.05,
and 0.01\M12, $k= 1$, $\zeta = 0.50$, and $\nu$ varying with the galactic mass.
The normalization mass is the present-day value of the total mass in 
alive stars  (see the data of Table 5)} 
\label{fig12}
\end{figure}

The metallicity distribution among the stars of the model galaxies of Table 5
is given in Figs.~\ref{fig11} and \ref{fig12}. In the former, we show the
relative total mass  ${\rm dM}$ 
in stars per metallicity bin. This is normalized to 
${\rm M_{L}(T_{G})}$. For the purposes of comparison we present also one case of
closed-box models (the 3\M12\ galaxy: thick line). 

The metallicity distribution in galaxies of different mass in shown is
Fig.~\ref{fig12} displaying the cumulative fractionary  mass in alive stars as
a function of ${\rm Z}$. 
In this case the distribution is normalized to the total
mass in alive stars at the age of 15 Gyr. We notice that while for the most
massive galaxy only 20\% of the stars have a metallicity below solar, this 
percentage increases at decreasing galaxy mass. Furthermore, the maximum
metallicity decreases at decreasing galaxy mass. Indeed, galaxies less massive
than 1\M12\ have only stars with ${\rm Z <Z_{\odot}}$. 

%%%%%%%%Figure 9 %%%%%%%%%%%
%%%%%%%%%%%%%%%%%%%%%%%%%%%

%%%%%%%%Figure 10 %%%%%%%%%%%

%%%%%%%%Figure 11 %%%%%%%%%%%

%%%%%%%%Figure 12 %%%%%%%%%%%

\subsection{Galactic winds}

The onset of galactic winds occurs within the first 0.5 Gyr for all the masses
under consideration (cf. the entries of Table 5). In contrast to previous
models of elliptical galaxies with galactic winds, in which the onset of these
is found to occur later in more massive objects (cf. BCF and references
therein), in these models the wind occurs earlier in more massive galaxies than
in the low mass ones. This can be understood as the result of the  efficiency
of star formation per unit mass of gas ($\nu$) increasing from $\nu=1$  to
$\nu=12$ as the galactic mass  goes from 0.01 to 3\M12. This trend is imposed
by the simultaneous fit of the slope and mean colours of the 
CMR and the dependence of the UV excess on the
galaxy luminosity and hence mass (see below). 
 
The result agrees with the suggestion by Matteucci (1994) to explain the high
value of the [Mg/Fe] observed in the brightest elliptical galaxies and its
decrease toward the solar value at decreasing luminosity of the galaxy. 

Finally, we like to remark that owing to the onset of galactic winds and
subsequent interruption of the star formation activity, the remnant galaxy made
of stars has a mass which is only a fraction of its nominal value. The real
current masses of the galaxies at the stage of the galactic wind are given in
Table~6. This contains the asymptotic mass \ML\ (column 1), the age at onset of
the galactic wind (column 2), the sum of the star and gas mass in units of
\M12\ (column 3), and finally the mass in stars (column 4) in the same units. 

\setcounter{table}{5}
%%%%%%%%%Table 6 (la massa in stelle ed in gas dei modelli) %%%%%
\begin{table}[htb]
\vskip 0.3 cm
%%%%%%%%%%%%%%%%%%%%%%%%%%%%%%%%%%%%%%%%%
\begin{center}
\vskip 0.2 cm
\caption{The mass of gas and stars at the onset of galactic winds for models
with different mass. The age ${\rm t_{gw}}$ at which the galactic wind  
occurs is in Gyr. Masses are in units of \M12. } 
\vskip 0.25 cm
\scriptsize
\begin{tabular} {c c c c}
\hline
\hline
 & & & \\
 \multicolumn{1}{c}{${\rm M_{L}(T_{GAL})}$              } &
 \multicolumn{1}{c}{${\rm t_{g\omega}}$                 } & 
 \multicolumn{1}{c}{${\rm M_{L}(T)S(t) + M_{L}(T)G(t) }$ } & 
 \multicolumn{1}{c}{${\rm M_{TOT}}$                     } \\
 & & &  \\
\hline
     &        &               &       \\
3    & 0.26   & 2.088 + 0.696 & 2.784 \\
1    & 0.31   & 0.648 + 0.312 & 0.960 \\
0.5  & 0.35   & 0.295 + 0.191 & 0.485 \\
0.1  & 0.34   & 0.041 + 0.056 & 0.097 \\
0.05 & 0.29   & 0.016 + 0.032 & 0.048 \\
0.01 & 0.43   & 0.002 + 0.008 & 0.010 \\
     &        &               &      \\
\hline
\hline
\end{tabular}
\end{center}
\label{tab6}
\end{table}

\subsection{Colour-magnitude relation}

\begin{figure}
%\picplace{ 9cm}
\psfig{file=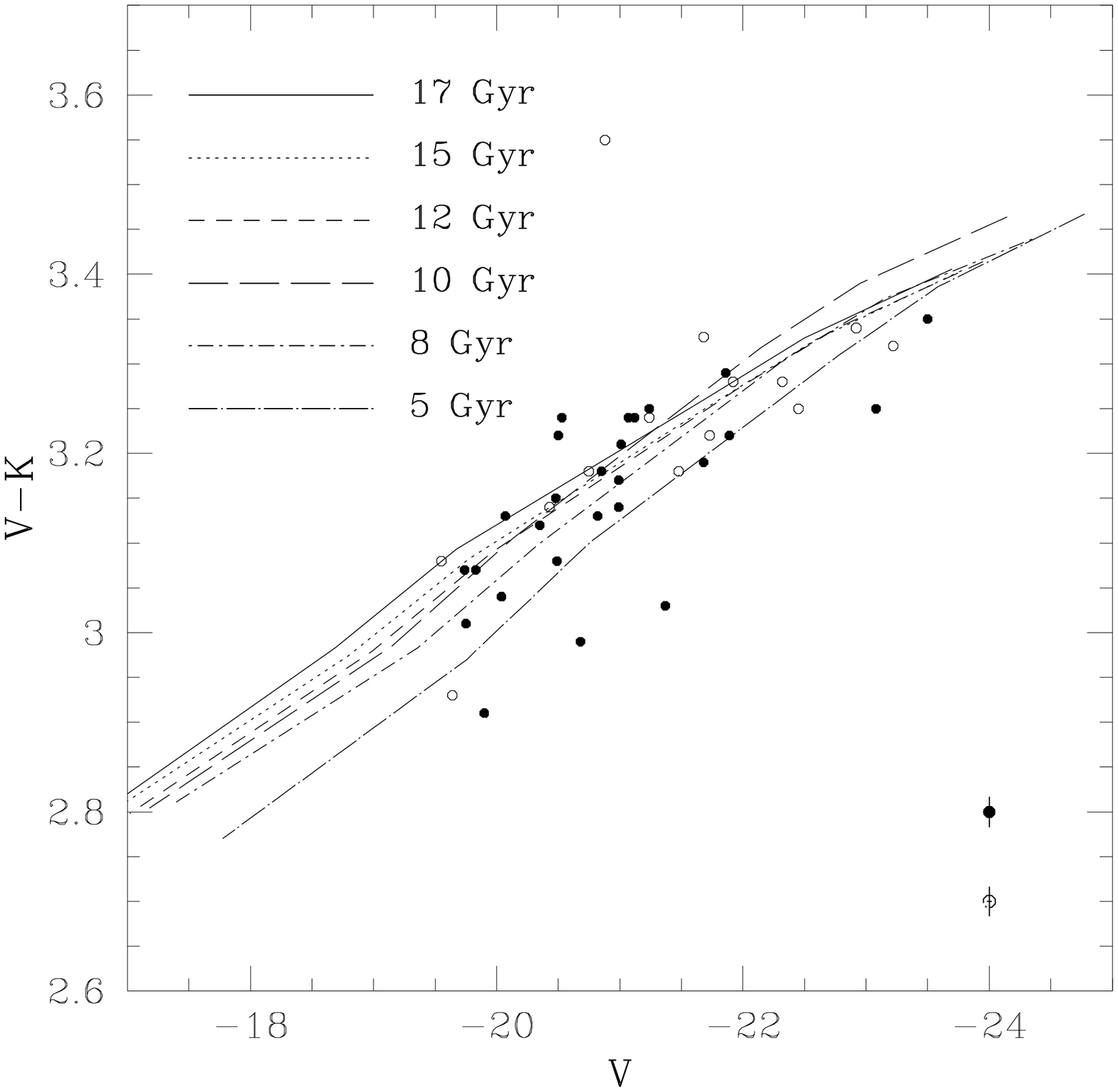,height=9.0truecm,width=8.5truecm}
\caption{The CMR for model galaxies with mass of 3, 1, 0.5, 0.1, 0.05, and 0.01
\M12 and the same set of parameters as in Fig.~\protect\ref{fig12}. The star
formation rate is characterized by $k = 1$. As indicated CMRs for different
ages are plotted. The data are for the galaxies in Virgo ({\em open circles})
and Coma ({\em filled circles}) by Bower et al. (1992a,b). The typical 
uncertainty in the photometric data is shown by the vertical bars }
\label{fig13}
\end{figure}

The CMR for the models of  Table 5 is compared with the observational one in
Fig.~\ref{fig13}. The theoretical CMR is given for several values of the age in
the range 5 to 17 Gyr. The data are from Bower et al. (1992a,b):  open circles
are the galaxies in Virgo, while the filled circles are the galaxies in Coma.
The absolute magnitudes ${\rm V}$ are calculated assuming for Virgo the distance
modulus ${\rm (m-M)_o=31.54}$ (Branch \& Tammann 1992) 
and applying to the galaxies in Coma the shift ${\rm \delta (m-M)_o=3.65}$ 
(Bower et al.\ 1992a,b). As expected the theoretical CMR nicely fit both the 
slope and the mean colours of the observational CMR.

%%%%%%%%Figure 13 %%%%%%%%%%%

\subsection{Mass to blue luminosity ratio}

\begin{figure}
%\picplace{ 9cm}
\psfig{file=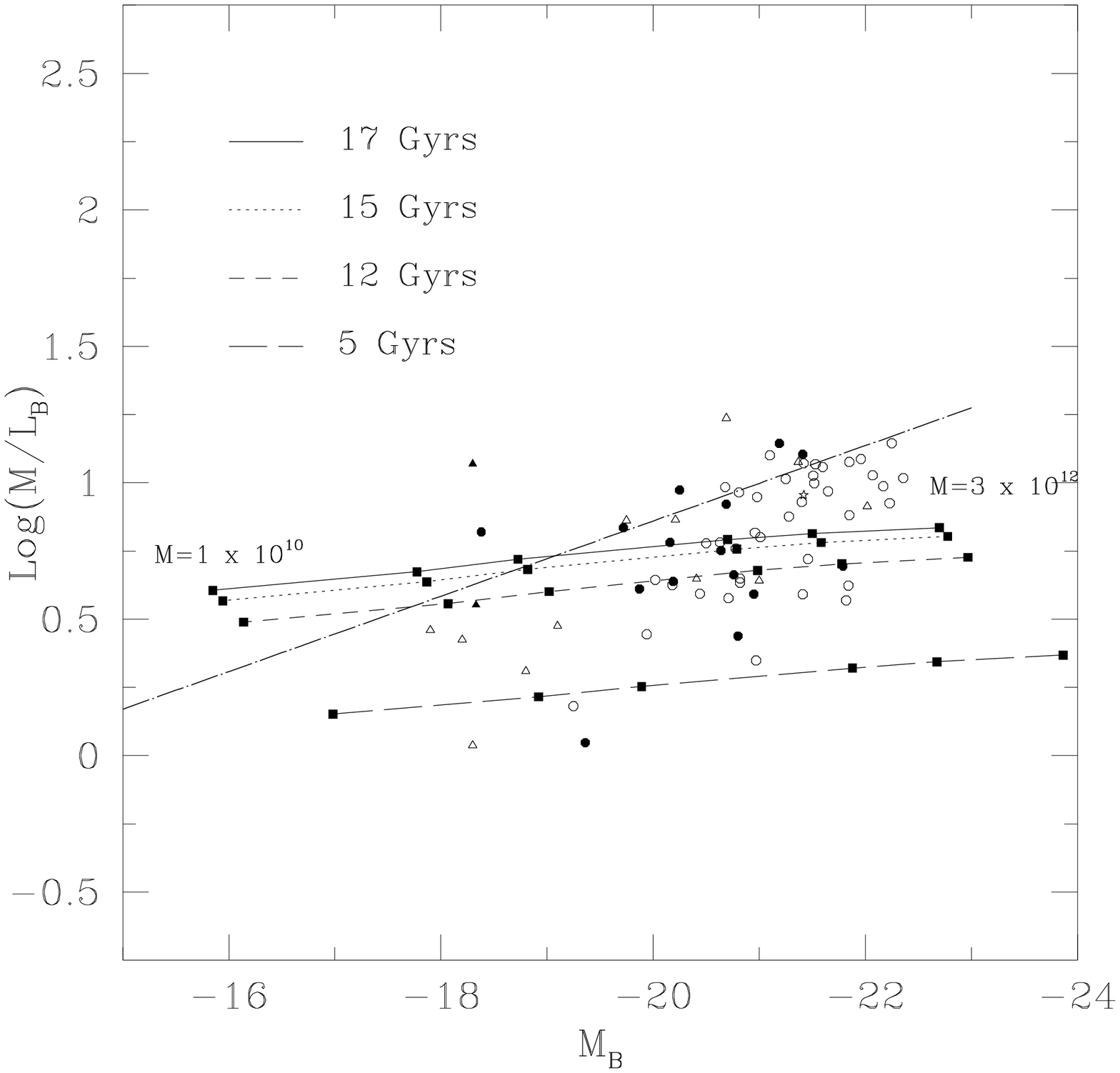,height=9.0truecm,width=8.5truecm}
\caption{The logarithm of the mass to $\rm B$-luminosity ratio 
${\rm M/L_{B}}$ versus the
absolute blue magnitude $\rm M_{B}$ for models with $k = 1$, 
$\nu$ increasing with the galactic mass, and
$\zeta = 0.50$. The mass used to calculate ${\rm M/L_{B}}$ and 
${\rm M_{B}}$ refers to the
present-day mass in the form of alive stars. The ratio 
${\rm M/L_{B}}$ is expressed
in solar units. The displayed quantities refer to ages form 17 to 5 Gyr. The
dashed line is the relation by Terlevich \& Boyle (1993) for ${\rm H_{0} = 50
Km s^{-1} Mpc^{-1} }$. The data are by Bender et al. (1992, 1993), i.e.
 open
dots: giant elliptical's; full dots: intermediate elliptical's; 
stars: bright dwarf elliptical's;  open squares: compact elliptical's;
open triangle: bulges        } 
\label{fig14}
\end{figure}

The mass to blue luminosity ratio ${\rm M/L_B}$ 
is found to be very sensitive to the
IMF,  and at given slope to $\zeta$. The observational ${\rm M/L_B}$ 
ratios (in solar
units) of Bender et al. (1992, 1993) and Terlevich \& Boyle (1993) --- scaled
to the Hubble constant ${\rm H_0=50~~km~~sec^{-1}~~Mpc^{-1} }$ --- are 
in the range
1 to 18. With the adopted IMF (eq. 17) the mean value of the observational data
is matched adopting $\zeta=0.50$. The theoretical ${\rm M/L_B}$ ratios  for the
models of Table 5 are given in Table~7, which lists the age in Gyr in column
(1), ${\rm M_{L}(T_{G})}$ 
in column (2), the blue magnitude $M_{B}$ in column (3),
(B--V) colour in column (4), the blue-luminosity  ${\rm L_{B}}$ 
in column (5), and
the mass to blue luminosity ratio ${\rm M/L_B}$ in column (6). 
The comparison with
the observational data is presented in Fig.~\ref{fig14} for different values of
the age. Although a certain spread among the ages of galaxies cannot be
excluded, the theoretical ${\rm M/L_B}$ versus ${\rm M_{B}}$ 
relationship seems to be
flatter than the mean slope traced by the observational data. The analysis of
the problem shows that in order to reconcile theory with observation either a
variation of the IMF slope or the fraction $\zeta$ with the galactic mass or
both should be  required (cf. Renzini \& Ciotti 1993). 

%%%%%%%%Figure 14 %%%%%%%%%%%

%\setcounter{table}{6}
%%%%%%%%%Table 7 (I rapporti massa luminosita`) %%%%%
\begin{table}[htb]
\vskip 0.3 cm
%%%%%%%%%%%%%%%%%%%%%%%%%%%%%%%%%%%%%%%%%
\begin{center}
\vskip 0.2 cm
\caption{The mass to blue-luminosity ratio (in solar units)
as function of the age for the model galaxies with different mass.  }
\vskip 0.25 cm
\scriptsize
\begin{tabular} {c c c c c c}
\hline
\hline
 & & & & & \\
 \multicolumn{1}{c}{${\rm Age}$} & 
 \multicolumn{1}{c}{${\rm M_{L}(T_{G})}$} &
 \multicolumn{1}{c}{${\rm M_{B}}$} & 
 \multicolumn{1}{c}{(B--V)} &
 \multicolumn{1}{c}{ $ {\rm \frac {L_{B} } {L_{\odot} }  } $} &
 \multicolumn{1}{c}{ $ {\rm \frac {M     } {L_{B}     }  } $} \\ 
 & & & & & \\
\hline
 & & & & & \\
 17 & 3.00 & -22.70 & 0.998 & 1.867e11 & 6.844 \\
 17 & 1.00 & -21.50 & 1.000 & 6.200e10 & 6.516 \\
 17 & 0.50 & -20.70 & 0.995 & 2.970e10 & 6.195 \\
 17 & 0.10 & -18.73 & 0.948 & 4.808e9  & 5.241 \\
 17 & 0.05 & -17.77 & 0.916 & 2.003e9  & 4.719 \\
 17 & 0.01 & -15.85 & 0.861 & 3.398e8  & 4.032 \\
 & & & & & \\
 15 & 3.00 & -22.78 & 1.001 & 2.008e11 & 6.364 \\
 15 & 1.00 & -21.58 & 1.002 & 6.699e10 & 6.031 \\
 15 & 0.50 & -20.79 & 0.998 & 3.218e10 & 5.718 \\
 15 & 0.10 & -18.82 & 0.955 & 5.238e9  & 4.811 \\
 15 & 0.05 & -17.87 & 0.928 & 2.180e9  & 4.335 \\
 15 & 0.01 & -15.94 & 0.880 & 3.712e8  & 3.691 \\
 & & & & & \\
 12 & 3.00 & -22.97 & 0.981 & 2.397e11 & 5.333 \\
 12 & 1.00 & -21.78 & 0.985 & 8.009e10 & 5.044 \\
 12 & 0.50 & -20.98 & 0.983 & 3.855e10 & 4.773 \\
 12 & 0.10 & -19.02 & 0.946 & 6.310e9  & 3.994 \\
 12 & 0.05 & -18.07 & 0.922 & 2.623e9  & 3.603 \\
 12 & 0.01 & -16.14 & 0.880 & 4.442e8  & 3.084 \\
 & & & & & \\
 10 & 3.00 & -23.18 & 0.967 & 2.903e11 & 4.403 \\
 10 & 1.00 & -21.98 & 0.973 & 9.612e10 & 4.203 \\
 10 & 0.50 & -21.18 & 0.971 & 4.600e10 & 4.000 \\
 10 & 0.10 & -19.20 & 0.932 & 7.447e9  & 3.384 \\
 10 & 0.05 & -18.24 & 0.908 & 3.087e9  & 3.061 \\
 10 & 0.01 & -16.31 & 0.868 & 5.214e8  & 2.627 \\
 & & & & & \\
 8  & 3.00 & -23.41 & 0.936 & 3.607e11 & 3.543 \\
 8  & 1.00 & -22.21 & 0.946 & 1.188e11 & 3.401 \\
 8  & 0.50 & -21.40 & 0.946 & 5.665e10 & 3.248 \\
 8  & 0.10 & -19.43 & 0.909 & 9.171e9  & 2.748 \\
 8  & 0.05 & -18.47 & 0.887 & 3.809e9  & 2.481 \\
 8  & 0.01 & -16.55 & 0.850 & 6.468e8  & 2.118 \\
 & & & & & \\
 5  & 3.00 & -23.87 & 0.910 & 5.475e11 & 2.334 \\
 5  & 1.00 & -22.68 & 0.908 & 1.830e11 & 2.208 \\
 5  & 0.50 & -21.88 & 0.902 & 8.790e10 & 2.093 \\
 5  & 0.10 & -19.89 & 0.862 & 1.407e9  & 1.791 \\
 5  & 0.05 & -18.92 & 0.837 & 5.754e9  & 1.642 \\
 5  & 0.01 & -16.98 & 0.792 & 9.656e8  & 1.419 \\
 & & & & & \\
\hline
\hline
\end{tabular}
\end{center}
\label{tab7}
\end{table}

\subsection{On the nature of the UV excess}

Key features of the UV excess (Burstein et al. 1988) to be interpreted by the
galactic models are: 

\begin{itemize}
\item{ The correlation of the \UVex\ colour with the index ${\rm Mg_2}$, 
the velocity
dispersion $\Sigma$, and the luminosity (mass) of the galaxy. } 

\item{ The drop-off of the UV  flux short-ward of about $1000 \AA$ observed in
the nucleus of M31 by Ferguson et al. (1991) and Ferguson \& Davidsen (1993)
which requires that  the temperature of the emitting source must be about
25,000 K.  Only a small percentage of the $912 \leq \lambda \leq 1200 $ \AA\
flux can be coming from stars hotter than 30,000 K and cooler than 20,000 K. } 
\end{itemize}

Excluding ongoing star formation,   the UV excess owes its origin to an old
component that gets hot enough to power the integrated spectral energy
distribution (ISED) of a galaxy in the far UV regions. Three candidates are
possible each of which likely contribute to the total emission in proportions
that depend on the particular history of star formation of the galaxy under
consideration (cf.  Greggio \& Renzini 1990, BCF)

%%%%%%%%Figure 15a %%%%%%%%%%%
\begin{figure}
%\picplace{ 9cm}
\psfig{file=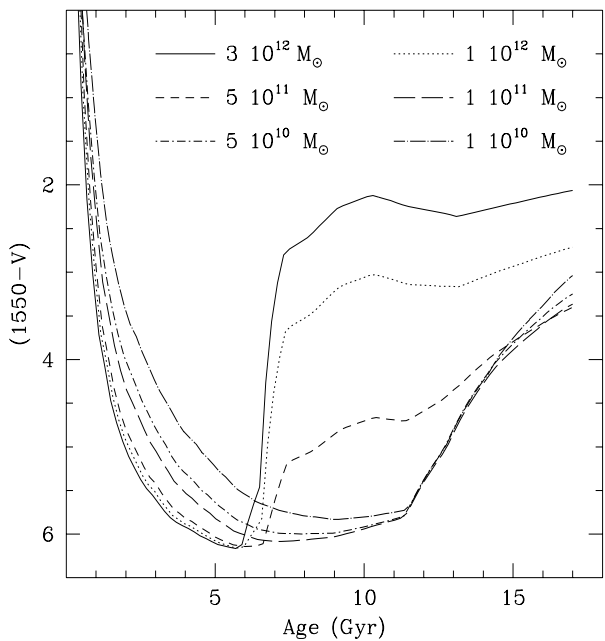,height=9.0truecm,width=8.5truecm}
\caption{ The colour (1550--V) as function of the age for galaxies of 
different \M12\ and mean and maximum metallicity. These latter increase with
the galactic mass. Note the gradual transition going from the 3\M12\ galaxy, 
in which H-HB and AGB-manqu\'e stars dominate the UV flux, to the 0.01 \M12\
galaxy, in which only the P-AGB stars contribute to it  } 
\label{fig15a}
\end{figure}

\begin{itemize}
\item{ The classical post asymptotic giant branch (P-AGB) stars (see Bruzual
1992, Bruzual \& Charlot 1993, Charlot \& Bruzual 1991). However, even if a
small drop-off can be predicted just long-ward of 912 \AA, they cannot be the
sole source of UV flux because of their high mean temperature (about 100,000 K)
and weak dependence on the metallicity. Nevertheless, these stars have been
detected with HST observations in the nucleus of M31 and can contribute by as
much as 50\% to the UV light (Bertola et al.  1995). } 

\item{ Very blue horizontal branch (HB) stars of extremely low metallicity (Lee
1994).  These stars have effective temperatures  hotter than about 15,000 K but
much cooler than those of the P-AGB stars. Therefore, depending on their actual
effective temperature, they can generate ISEDs in agreement with the
observational data. However, most likely they are not  the dominant source of
the UV flux because BCF showed that  in the wavelength interval $2000 < \lambda
< 3000$ \AA\ the ISEDs of the  bulge of M31 and of elliptical galaxies like
NGC~4649 are fully consistent with the notion that virtually  no stars with
metallicity lower than Z=0.008 exist in the mix of stellar populations. } 

\item{  Finally, the  hot horizontal branch (H-HB)  and AGB-manqu\'e stars of
very high metallicity (say Z$>>$0.07) which are expected to be present albeit
in small percentages in the stellar content of  bulges and elliptical galaxies
in general. Indeed, these stars have effective temperatures in the right
interval and generate ISEDs whose intensity drops short-ward of  about 1000
\AA\ by the amount indicated by the observational data (Ferguson et al. 1991,
Ferguson \& Davidsen 1993).  } 
\end{itemize}

The appearance of the various types of UV sources is governed by several
important physical factors, each of which is affected by a certain degree of
uncertainty still far from being fully assessed. These are the efficiency of
mass loss during the RGB and AGB phases, in particular whether or not it
contains a strong dependence on the metallicity, the enrichment law \dydz\
fixing the correspondence between helium and metal contents of the stellar
models, and finally for the specific case of P-AGB stars the detailed relation
between the initial and final mass of the stars at the end of the AGB phase. It
is beyond the scope of this paper to examine  the above uncertainties in
detail. They have amply been discussed by Greggio \& Renzini (1990) and BCF to
whom we refer. Suffice to recall that 

\begin{itemize}
\item{ P-AGB stars are always present in the stellar mix of a galaxy. The major
difficulty with these stars is their high effective temperature, and the
precise relation between their mass and that of the progenitor (and hence the
turn-off mass and  age).  The response of the UV flux to details of
initial-final mass relation (for instance its dependence on the metallicity) is
so strong that firm conclusions cannot yet be reached. For instance at ages
older than about 10 Gyr, the whole problem is driven by the initial-final mass
relation in the mass range 0.8$\div$1.0 \Msun. } 

\item{ H-HB and AGB-manqu\'e are  thought to be typical of low mass stars with
metallicity ${\rm Z >> Z_{\odot}}$. 
Their formation can be obtained either with low
values of the enrichment ratio ($\dydz \simeq 1$) and strong dependences of the
mass-loss rates on the metallicity (Greggio \& Renzini 1990) or even with
canonical mass-loss rates and suitable enrichment laws ($\dydz \simeq 2.5$) as
in  Horch et al. (1992) and Fagotto et al. (1994a,b,c). } 
\end{itemize}

The present spectro-photometric models owe their UV flux to the combined effect
of P-AGB stars and H-HB and AGB-manqu\'e objects, these latter being very
sensitive to the  metallicity distribution. The different nature of the
UV source at varying galactic mass and mean and maximum metallicity in turn
 is best 
illustrated by the age-(1550-V) colour relation  shown in Fig.~\ref{fig15a}. 
In the most massive objects the UV flux 
is dominated by the H-HB and AGB-manqu\'e stars (uprise in the age range
5.6 to 6.8 Gyr), whereas in the less massive galaxies the UV excess is
entirely produced by the P-AGB stars (gradual uprise starting from 11.5 
Gyr). 

%%%%%%%%Figure 16 %%%%%%%%%%%
\begin{figure}
%\picplace{ 9cm}
\psfig{file=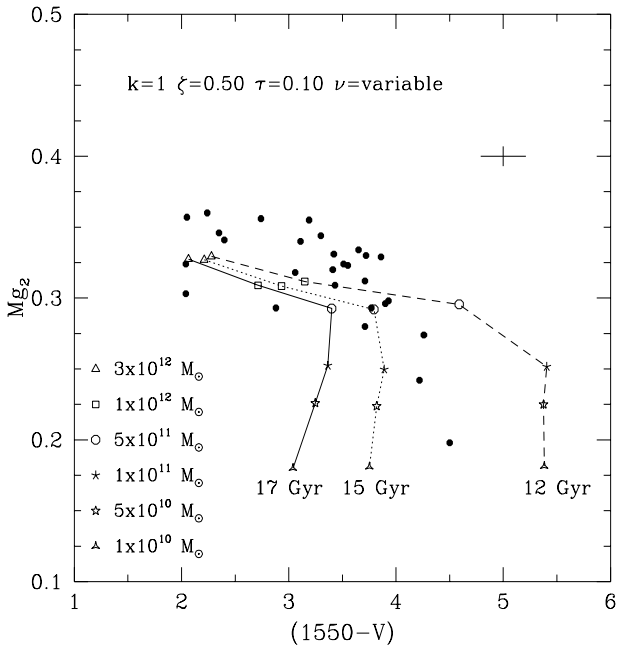,height=9.0truecm,width=8.5truecm}
\caption{Comparison of the theoretical relation ${\rm Mg_{2}}$ versus 
(1550--V) relation with the
data by Burstein et al.\ (1988) indicated by the  filled circles. 
The theoretical ${\rm Mg_{2}}$
is obtained from the (V--K) colour according to the analytical fit due to
Buzzoni et al. (1992): ${\rm Mg_{2} = 0.24(V-K) - 0.49}$. 
The effect of the age
is brought into evidence by drawing with different symbols the 
${\rm Mg_{2}}$-(1550--V) relation for three different values of the age, i.e.
12, 15, and 17 Gyr. Along the three curves (dashed, dotted and 
solid, respectively,  the galaxy mass and maximum metallicity decrease 
from left to right.  The data are from Table 5. 
The cross shows the typical error bars of the data. See the text 
for more details} 
\label{fig16}
\end{figure}

\subsection{ The  $ Mg_2$ index }

The correlation between the colour (1550--V) and the $Mg_2$ index is shown in
Fig.~\ref{fig16} together with the observational data of Burstein et al.
(1988). The index ${\rm Mg_2}$ is derived from  
the broad band colour (V--K) using
the calibration of Buzzoni et al. (1992), which fits the data for elliptical
galaxies by Davies et al. (1993) and Frogel et al. (1978) and  is also
consistent with those between (V--K) and central velocity dispersion
(${\rm Log\Sigma_0}$) by Bower et al. (1992a,b) and between 
${\rm Mg_2}$ and ${\rm Log\Sigma_0}$
by Bender et al. (1992, 1993). 
This choice is imposed by the lack of sufficient
resolution in the library of stellar spectra that does not allow us to apply the
standard definition of this narrow band feature (Faber et al. 1977). Detailed
calculations of line strength indices according to the Worthey et al. (1994)
calibrations  are presented in Bressan et al. (1995) to whom we refer. 
In spite
of this limitation, the ${\rm Mg_2}$ versus (1550--V) relationship of
Fig.~\ref{fig16} is fully adequate to  the present purposes.  As expected the
${\rm Mg_2}$ versus (1550--V) relation 
depends both on the age and metallicity via the galactic mass thus
partly explaining the scatter in the observational data. We notice 
in particular that at given age the ${\rm Mg_2}$ index decreases at 
decreasing galactic mass (mean and maximum metallicity in turn). 
There is a marginal discrepancy in the theoretical ${\rm Mg_2}$ indices that
 are somewhat smaller than the observational values. This can be attributed to 
the effect of the $\alpha$-element partition 
discussed in section 4.6. Correcting for it would likely increase the
theoretical ${\rm Mg_2}$ by 0.05 thus bringing the theoretical prediction to
agree with the observational data.

\subsection{ Detailed comparison of ISEDs  }
\begin{figure}
%\picplace{ 12cm}
\psfig{file=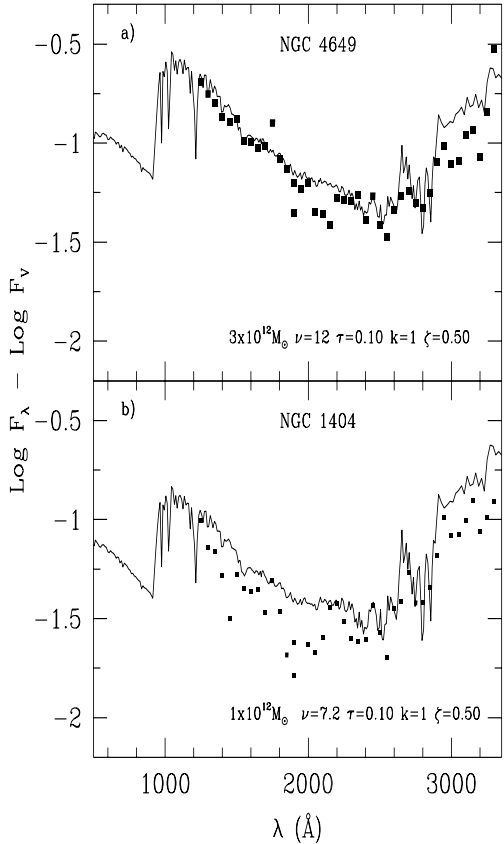,height=12.0truecm,width=8.5truecm}
\caption{{\bf Panel (a)}: comparison of the spectrum of NGC~4649 (filled
squares) with the theoretical ISED of the model with mass 3\M12\ (solid line).
The observational ISED is from Burstein et al. (1988). Note the remarkable
agreement with the data. {\bf Panel (b)}: the same as in Panel (a)
 but for the galaxy
NGC~1404 (filled squares) and the theoretical ISED of the model with mass
1\M12.} 
\label{fig15}
\end{figure}

The theoretical ISEDs reasonably match the observational ones both in the 
far UV and all over a large range of wavelength from the UV to
the near infrared. To illustrate the first point we show how 
 our models match the observed ultraviolet ISED of two prototype
galaxies: NGC~4649 a strong source with (1550--V)=2.24 
and NGC~1404 an intermediate source with
(1550--V)=3.30. The former is best matched by the
model with 3\M12\ and age of 15 Gyr,  whereas the latter is reasonably 
reproduced  by the model with 1\M12\ and same
age. The comparison is shown in the two panels of  Fig.~\ref{fig15}. Firstly,
we notice that there is no excess of flux in the range 2000 - 3500 \AA\ as it
occurred with the closed-box approximation (see BCF). As expected, the infall
models are able to solve the analog of the G-Dwarf problem in the ISED of
elliptical galaxies.  
Secondly, the source of  UV flux gets more and more
weighted toward  the H-HB and AGB manqu\'e channel as compared to the classical
P-AGB stars at increasing galaxy mass and hence maximum attainable 
metallicity. The trend can be inferred from the
decreasing flux intensity at about 1000 \AA\ with respect to that at shorter
wavelengths passing from a massive to a less massive galaxy, and also from the
relative value of the flux in the region 1000 - 2000 \AA\ to that in the visual
(compare top and bottom panels of Fig.~\ref{fig15}). Furthermore, the
theoretical ISEDs have the kind of fall-off at about 1000 \AA\ indicated by the
observations of Ferguson et al. (1991) and Ferguson \& Davidsen (1993) in
elliptical galaxies with UV excess.  Finally, the two model galaxies have
(1550--V) colours  that fairly agree with the observational data: 2.21 for the
most massive galaxy and 2.93 for the less massive one (see the data in Table
5). 

%%%%%%%%Figure 15 %%%%%%%%%%%

\begin{figure}
%\picplace{ 12cm}
\psfig{file=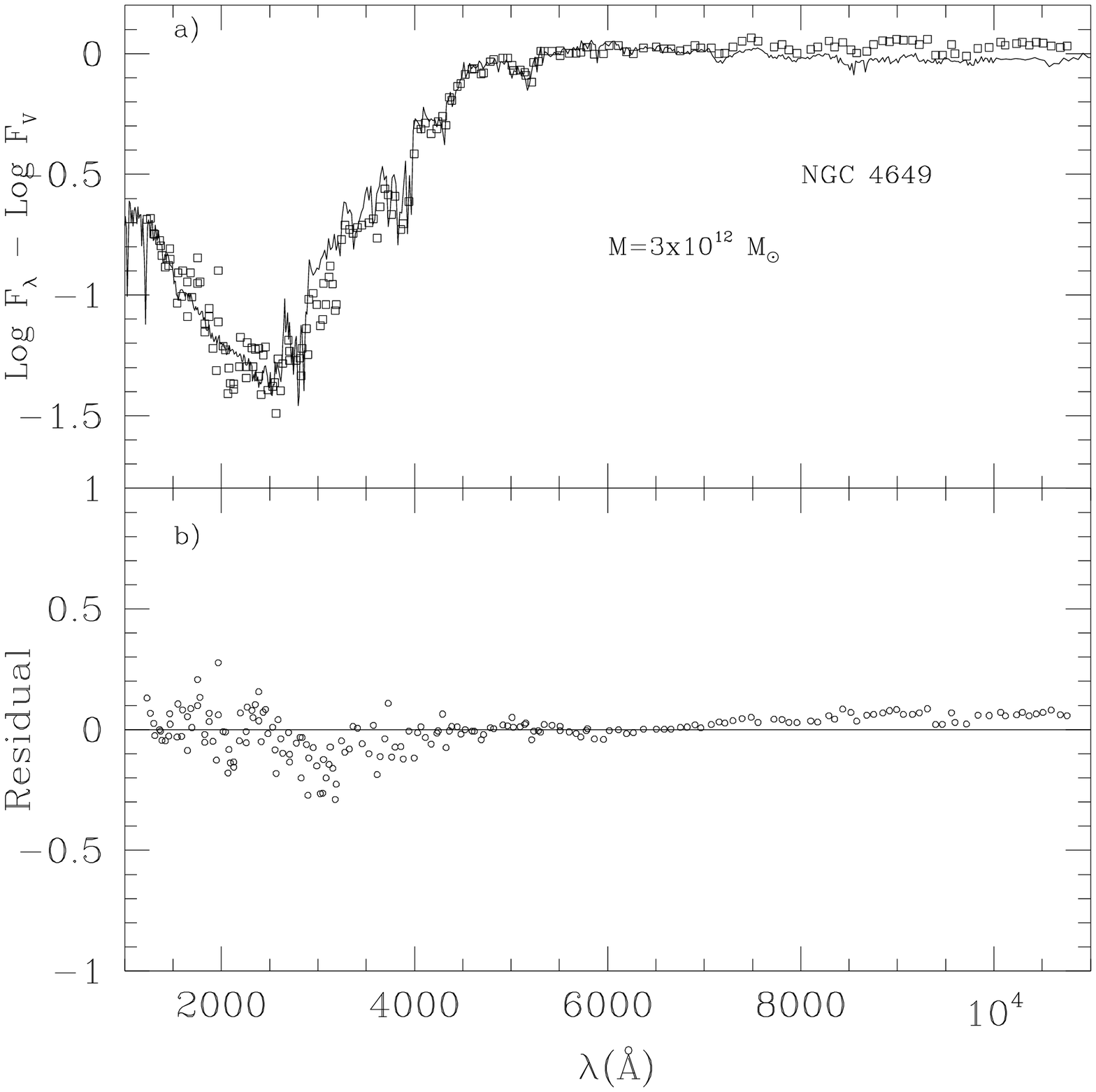,height=12.0truecm,width=8.5truecm}
\caption{{\bf Panel (a)}: 
the observed spectrum of the elliptical galaxy NGC~4649
indicated by open squares (Lucio  Buson, 1993, private communication). The
solid line shows the theoretical ISED for  the model with 3 \M12, $k = 1$, 
$\nu=12.0$, $\zeta=0.50$, $\tau=0.10$ Gyr, and age of 15 Gyr. Note the good
agreement. {\bf Panel (b)}: residuals 
(open circle) between the theoretical ISED and
the data.} 
\label{fig17}
\end{figure}

Concerning the second point, we  compare the ISED of NGC~4649 from
1000 to 11,000 \AA\ (from the far UV to the near infrared) with the theoretical
ISED of the 3\M12\ model. 
The comparison is shown in Fig.~\ref{fig17}. Mounting
of the theoretical ISED on the observational one is made imposing coincidence
at about 5500 \AA. In general the agreement is satisfactory. However there are
a few points of discrepancy worth being noticed. Indeed, there is a deficiency
of flux in the theoretical ISED  above 7500 \AA\ and some scatter in the UV
region as indicated by the residuals shown in panel b of Fig.~\ref{fig17}. 

%%%%%%%%Figure 17 %%%%%%%%%%%

%%%%%%%%Figure 18 %%%%%%%%%%%
\begin{figure}
%\picplace{ 9cm}
\psfig{file=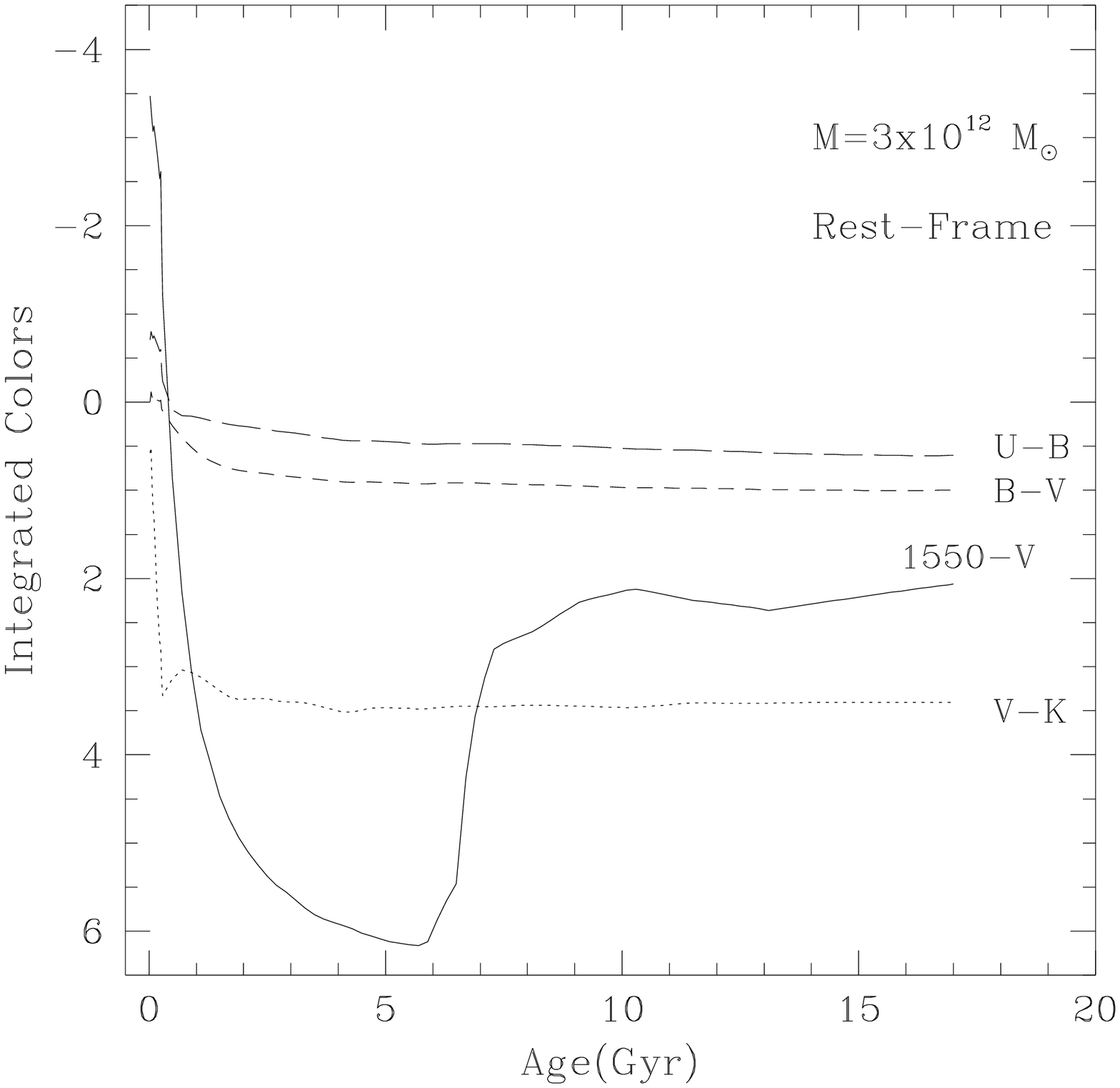,height=9.0truecm,width=8.5truecm}
\caption{The integrated colours (1550--V), (U--B), (B--V) and (V--K) as a
function of the rest frame age for the model  with  3 \M12, $k= 1$, $\nu =
12.0$, $\zeta = 0.50$ and $\tau = 0.10$ Gyr} 
\label{fig18}
\end{figure}

\subsection{UV-excess as  age probe}

In this section, we discuss  the possibility of dating galaxies by means of the
UV-excess. BCF have shown that signatures in the colour evolution (like those
caused by the appearance of AGB stars) manifesting themselves  at early epochs
cannot be used  as age probes  because their detection as a function of the
red-shift is wiped out by cosmological effects. On the contrary, the onset of
the UV-excess caused by the old, high metallicity stars is a very promising
candidate.  The most favourable object to look at is a massive (luminous)
object with strong UV-excess, such as our model with 3\M12. 
 
To illustrate the point  in Fig.~\ref{fig18} we show the evolution of the
colours, (U--B), (B--V), (V--K) and (1550--V) as a function of the age for the
3\M12\ model. 

Following the initial period of stellar activity (in this galaxy confined
within the first 0.26 Gyr), no sign of evolution is visible in the colours
(U--B) and (B--V) which get redder and redder at increasing age. The colour
(V--K) has a marked drop-off at the age at which the first AGB stars occur (a
few $10^8$ yr) and remains almost constant afterward. The very young age at
which the drop-off occurs is the cause of the failure of the (V--K) colour as
age indicator. As a matter of facts, this young age would correspond to high
values of the red-shift thus rendering detection difficult and cosmological
distortion of the ISED and hence colour very large. 

The colour (1550--V) is by far more suited. First,  following the interruption
of the star formation activity by galactic winds, it colour drops to very ``red
values'' (low UV flux). Subsequently, at much older ages in coincidence with
the appearance of the old stars emitting UV radiation (P-AGB and above all H-HB
and AGB-manqu\'e objects) it becomes ``blue'' again (see Fig.~\ref{fig15a}). 
For our test galaxy 
the uprise in the (1550--V) colour occurs at the age of about 5.6 Gyr to
which a relatively low value of the red-shift corresponds. 
 
It goes without saying that the age in question depends the adopted
${\Delta}Y/{\Delta}Z$ and efficiency of mass loss during the RGB and AGB
phases. The explanation of the UV-excess in elliptical galaxies imposes some
constraints both on ${\Delta}Y/{\Delta}Z$ and mass loss. Our choices for both
constitute a sort of lower limit, and therefore the corresponding age of the
UV-upturn is a sort of upper limit. However, much higher ${\Delta}Y/{\Delta}Z$
and mass loss rates are not very likely. 

It follows from this analysis that not only the UV-excess is a powerful probe
of galaxy ages but the detection of its onset at a certain red-shift together
with the measurements of its fall off as a function of the red-shift would
allow us to establish  useful relations between ${\rm H_0}$, 
${\rm q_0}$ and ${\rm z_{for}}$.
Future observations will certainly make feasible the   detection of this
important evolutionary effect. 

\section{ Summary and conclusions }
In this paper we have presented one zone chemo-spectro-photometric  models of
elliptical galaxies with infall of primordial gas. Galaxies are supposed to be
made of two components, i.e. luminous and dark matter. While the mass of the
dark component is assumed to be 
constant with time, that of the luminous material
is supposed to fall at a suitable rate into the common potential well. 

The main motivation for this type of model was to mimic the collapse of a
galaxy and to cope with the difficulty reported by BCF using the closed-box
scheme, i.e. the analog of the G-Dwarf problem. The adoption of the infall
scheme adds another dimension to the problem, i.e. the time scale of mass
accretion $\tau$. Taking the free-fall time scale as a reference, instead of
changing $\tau$ with the galaxy mass, low mass galaxies would have got closer
and closer to the closed-box approximation, we prefer to adopt the unique value
 $\tau=0.1$ Gyr, i.e. the free-fall time scale for a typical galaxy with mass
0.5 \M12. 

The models also include the occurrence of galactic winds that have long been
considered the key physical process governing the CMR of elliptical galaxies.
The galactic winds are supposed to occur when the energy input from supernovae
and stellar winds from luminous early type stars equals the gravitational
binding energy of the gas. It goes without saying that this is a point of great
uncertainty of the models. In this respect our approach does not go beyond
other current models of elliptical galaxies with galactic winds taken into
account. Both the rate of supernova explosions and cooling of the energy
injected by these stars are treated in the standard fashion (see for instance
Arimoto \& Yoshii 1987, Matteucci \& Tornamb\'e 1987, Brocato et al.\ 1990,
Angeletti \& Giannone 1990). However, in agreement with BCF and in contrast
with other authors, we confirm that if the CMR of elliptical galaxies has to be
understood a mass-metallicity sequence of nearly coeval objects, in addition to
supernova explosions another source of energy must be at work in order to
trigger galactic winds before the metallicity has grown to such high values
that only very red colours are possible. 

The adoption of the infall scheme has the advantage over the closed-box
description that with the law of star formation proportional to 
${\rm M_{g}^{k}(t)}$,
the rate of star formation as a function of time starts small, grows to a
maximum and then declines thus easily avoiding the excess of zero metal stars
typical of the closed-box scheme. Indeed, infall models have the metallicity
distribution of its stellar content skewed towards relatively high
metallicities. 

As repeatedly said, the bottom line to contrive the various parameters of the
models has been the simultaneous fit of many properties of elliptical galaxies.
We have found that with the new library of stellar spectra in usage the
simultaneous fit of the (V--K) and (1550--V) colours is not possible  with the
closed-box description, whereas this is feasible with the infall scheme. The
fit of the slope and mean colours of the
 CMR and UV excess of elliptical galaxies required that the
efficiency of star formation per unit mass of gas ($\nu$) must increase with
the galaxy mass. As a consequence of this, the overall duration of the star
forming activity (before the onset of galactic winds)  in massive galaxies is
shorter  than in the low mass ones. This kind of trend is in agreement with the
hint coming from the observational dependence of the [Mg/Fe] ratio with the
galaxy luminosity (and hence mass) and the suggestion based on chemical models
advanced by Matteucci (1994). 

As far as the simultaneous fit of the CMR, (V--K) versus V,  and the (1550--V)
versus $Mg_{2}$ relation by our model galaxies as a whole imply that the two
observational sets of data refer to the same region of the galaxies. While the
data on the UV-excess, colour (1550--V) and $Mg_{2}$ index essentially refer to
the central parts of the galaxies (see Burstein  et al. 1988), the integrated
magnitudes and colours, such as V, (B--V), (V--K) etc., refer to the whole
galaxy. This is particularly true with the data we are using for galaxies in
the Virgo and Coma cluster. The presence of spatial gradients of colours and
metallicity that are known to exist in elliptical galaxies (Carollo et al.
1993, Carollo \& Danziger 1994, Davies et al.1993, Shombert et al. 1993)
indicates that the one-zone models (either closed or with infall) ought to
abandoned in favor of models in which the spatial distribution of mass density
and star formation rate is taken into account. It seem reasonable to argue that
higher metallicities can be present in the central regions thus facilitating
the formation of the right type of stellar source of the UV radiation (H-HB and
AGB-manqu\'e stars) whereas lower metallicities across the remaining parts of
the galaxies would make it easier to fit the others integrated colours such as
(V--K). Preliminary models of elliptical galaxies with spatial distribution of
total mass and gas density, star formation rates, metallicities and colours are
in progress (Tantalo 1994). 

An alternative scenario is been proposed by Elbaz et al. (1995) aimed at 
explaining the amount of iron present in the intra-cluster medium.
The goal is achieved by means of models for the evolution of 
elliptical galaxies with bimodal star formation. In brief,
most of the heavy elements are supposed to be produced by SNII during
a first violent phase during which only high-mass stars are formed.
This is followed by a phase of quiescence caused by 
galactic winds. Finally, star formation is supposed to start again with
normal IMF when sufficient gas has been made available by previously borne 
stars. However, using the Elbaz et al. (1994) models 
Gibson (1995) has tried to reproduce the present-day photometric and
chemical properties of elliptical galaxies with the conclusion 
that these models while successfully reproducing 
the observational data for the intra-cluster medium, fail
to reproduce the chemo-spectro-photometric properties of the present day 
elliptical galaxies.

\vspace{1truecm}

\noindent
{\bf Acknowledgements} \\
This work has been financially supported by the Italian Ministry of University,
Scientific Research and Technology (MURST), the National Council of Research
(CNR-GNA), and the Italian Space Agency (ASI).

\newpage
\newpage

\end{document}